\documentclass[letterpaper,12pt]{article}
\usepackage[margin=25mm]{geometry}
\usepackage{amsmath}
\usepackage{amsfonts}
\usepackage{amssymb}
\usepackage{graphicx}
\usepackage{verbatim}
\usepackage{subfigure}
\usepackage{caption}
\usepackage{setspace}

\usepackage{color,bm}
\usepackage{graphicx}
\usepackage{array,multirow}
\usepackage{float}
\usepackage{afterpage}
\usepackage{arydshln}
\usepackage{titlesec}
\usepackage{blkarray}
\usepackage{mathtools} 
\usepackage{pgfplots}
\usepackage{tikz}
\usepackage{multirow}
\usetikzlibrary{bayesnet}
\usepackage{blkarray}
\usepackage{multirow}
\usepackage[ruled,vlined]{algorithm2e}
\usepackage{booktabs}
\usepackage{enumitem}

\usetikzlibrary{shapes,arrows}

\usepackage{xr}

\usepackage{hyperref} 
\hypersetup{
	colorlinks=true,       
	linkcolor=blue,        
	citecolor=blue,        
	filecolor=magenta,     
	urlcolor=blue         
}
\usepackage{setspace}

\newtheorem{theorem}{Theorem}

\newtheorem{corollary}{Corollary}

\newtheorem{definition}{Definition}
\newtheorem{example}{Example}

\newtheorem{proposition}{Proposition}
\newtheorem{remark}{Remark}

\newenvironment{proof}[1][Proof]{\noindent\textbf{#1.} }{\ \rule{0.5em}{0.5em}}

\tikzstyle{place}=[circle,draw=black,fill=white,inner sep=0pt,minimum size=6mm]
\tikzstyle{group1}=[rounded corners,fill=blue!5,inner sep=1ex]

\tikzset{
  LabelStyle/.style = { rectangle, rounded corners, draw,
                        minimum width = 2em, fill = yellow!50,
                        text = red, font = \bfseries },
  VertexStyle/.append style = { inner sep=2pt,
                                font = \bfseries},
  EdgeStyle/.append style = {->,  left} }

\immediate\write18{texcount -tex -sum  \jobname.tex > \jobname.wordcount.tex}

\usepackage[sort]{natbib}

\title{Hypothesis Testing for Hierarchical Structures in Cognitive Diagnosis Models}
 \author{Chenchen Ma and Gongjun Xu \\ University of Michigan }
\date{} 

\pgfplotsset{compat=1.17}

\begin{document}
\maketitle

\pagestyle{plain}
\setcounter{page}{1}
\pagenumbering{arabic}

\begin{abstract}
Cognitive Diagnosis Models (CDMs) are a special family of discrete latent variable models widely used in educational, psychological and social sciences. In many applications of CDMs, certain hierarchical structures among the latent attributes are assumed by researchers to characterize their dependence structure. Specifically, a directed acyclic graph is used to specify hierarchical constraints on the allowable configurations of the discrete latent attributes. In this paper, we consider the important yet unaddressed problem of testing the existence of latent hierarchical structures in CDMs. We first introduce the concept of testability of hierarchical structures in CDMs and present sufficient conditions.
Then we study the asymptotic behaviors of the likelihood ratio test (LRT) statistic, which is widely used for testing nested models. Due to the irregularity of the problem, the asymptotic distribution of LRT becomes nonstandard and tends to provide unsatisfactory finite sample performance under practical conditions. We provide statistical insights on such failures, and propose to use parametric bootstrap to perform the testing. We also demonstrate the effectiveness and superiority of parametric bootstrap for testing the latent hierarchies over non-parametric bootstrap and the na\"ive Chi-squared test through comprehensive simulations and an educational assessment dataset. 

\end{abstract}

\section{Introduction}\label{chap1-sec-intro}
Cognitive Diagnosis Models (CDMs) are a popular family of discrete latent variable models that have been widely used in social and biological sciences.
In CDMs, a set of latent attributes are assumed to exist that can explain or govern the observed variables.
Under the framework of CDMs, one can get fine-grained inference on subjects' latent attribute profiles based on their multivariate observed responses, which further can be used to infer the subgroups of the population.
In practice, the discrete latent attributes often have special scientific meanings.
For example, in educational assessments, the latent attributes are assumed to be mastery or deficiency of target skills \citep{junker2001cognitive, de2011generalized};
in psychiatric diagnosis, they are modeled as presence or absence of some underlying mental disorders \citep{templin2006measurement, de2018analysis};
and in epidemiological and medical measurement studies, the latent attributes are interpreted as existence or nonexistence of some disease pathogens \citep{wu2016nested, wu2016partially, o2019causes}.

Various CDMs have been developed in the literature by modeling the interactions between the observed variables and the latent attributes differently under different cognitive diagnostic assumptions.
The Deterministic Input Noisy Output “AND” gate \citep[DINA;][]{haertel1989} model and the Deterministic Input Noisy Output “OR” gate \citep[DINO;][]{templin2006measurement} model are among the most basic ones. 
More general CDMs include the Generalized DINA (GDINA) model \citep{de2011generalized}, the General Diagnostic Model \citep[GDM;][]{von2005general}, the reduced Reparameterized Unified Model \citep[reduced-RUM;][]{dibello1995unified}, and the Log-linear Cognitive Diagnosis Models \citep[LCDM;][]{henson2009defining}.

Recently the relationship among the latent attributes, especially hierarchical structures among them, has gained increasing research interests.
For example, in a learning context, prerequisite relations are often postulated among the latent skill/concept attributes, which can be used to characterize the learning trajectory of the students \citep{dahlgren2006senior, jimoyiannis2001computer, simon2004explicating, wang2011using}.
Based on the prerequisite relationships among the latent attributes, learning materials or test items can be designed, and recommendations or remedy strategies can be generated accordingly.
The Attribute Hierarchy Model \citep{leighton2004attribute} is a variation of Tatsuoka's rule-space approach \citep{tatsuoka1983rule} and uses an adjacency matrix to explicitly defined the hierarchical attribute structures. 
\cite{templin2014hierarchical} proposed the Hierarchical Cognitive Diagnosis Models (HCDMs), which enforce hard constraints on hierarchical configurations of the latent attributes.
In particular, a Directed Acyclic Graph (DAG) was used to characterize the hierarchical dependence structure among the latent attributes, and impose hard constraints on possible latent attribute profiles under hierarchies.

Despite the popularity of CDMs, statistical inference of relationships among the latent attributes especially hierarchical structures has been a challenging yet unaddressed problem in the literature. 
For example, hypothesis testing plays an important role in validating the presence of suspected attribute hierarchies, which can provide guidance to practitioners for experiment design or data modeling \citep{templin2014hierarchical}.
However, to our best knowledge, there are no systematical testing procedures or statistical theories on hypothesis testing of latent hierarchical structures.  
Two natural questions about such testings are (1) when the hierarchical structures are testable and (2) how to conduct the hypothesis testing.
On the one hand, under the framework of CDMs, if the hierarchical structure under the null hypothesis cannot be distinguished from those under the alternative, we cannot test such a hierarchical structure and therefore it is untestable.
In fact, the testability of hierarchical structures is closely related to the identifiability of the models.
On the other hand, under the hierarchical constraints, the problem of testing latent hierarchical structures is equivalent to testing the sparsity structure of the set of latent attribute profiles in the population, 
that is, the sparsity structure of the population proportion parameter vector.
However, due to the identifiability and the irregularity issue that the true proportion parameters are on the boundary of the parameter space under hierarchical structures, the conventional asymptotic Chi-squared distribution may not hold for the likelihood ratio test.

Non-regularity issues of the likelihood ratio test are known to exist in many latent variable models such as finite mixture models, factor analysis, structural equation models, and random effects models \citep{chen2017finite, chen2020note}.
In particular, testing the sparsity structure of the proportion parameter vector in CDMs is closely related to the problem of testing the number of components in finite mixture models and latent class models \citep{nylund2007deciding, chen2017finite}.
However, testing the hierarchical structures in CDMs is even more challenging,  since it tests whether a specific set of the proportion parameters specified by the hierarchical structure under the null hypothesis is zero; 
and such a problem is further complicated due to the restrictions imposed by the structural $Q$-matrix and the discrete nature of the latent variables in CDMs.

In this paper, we focus on the problem of hypothesis testing for latent hierarchical structures. 
We first discuss the testability of latent hierarchical structures and present sufficient conditions under which hierarchical structures are testable in CDMs. 
Then under such conditions, we examine the asymptotic behaviors of the popularly used likelihood ratio test. 
Since the true proportion parameter is on the boundary of the parameter space, the asymptotic distribution of LRT becomes nonstandard  due to the lack of regularity \citep{self1987asymptotic}.
Moreover, the nonstandard limiting distribution of LRT is observed to not provide satisfactory finite-sample results under practical settings, and we provide statistical insights on such failures.
Specifically, we find that when the number of items is large or the item parameters are close to the boundary, the convergence of the nonstandard limiting distribution can be very slow and the test tends to fail.
Therefore we do not recommend using the nonstandard limiting distribution to conduct the hypothesis testing in practice.
Instead, based on these findings, we propose to use resampling-based methods to test hierarchical structures.
We conduct comprehensive simulations and comparisons between parametric bootstrap and nonparametric bootstrap and recommend using parametric bootstrap for testing latent hierarchies in CDMs.

The rest of the paper is organized as follows:
the model setup of hierarchical CDMs and the motivations of the problem are introduced in Section \ref{sec-model}. 
Sufficient conditions for testability of hierarchical structures and several illustrative examples are provided in Section \ref{sec-test}.
Studies on the likelihood ratio test and numerical results are presented in Section \ref{sec-LRT}.
Specifically, section \ref{sec-limiting} studies the asymptotic behaviors of LRT and provides insights on its failures in some situations.
Section \ref{sec-boot} presents simulation studies that compare parametric bootstrap and nonparametric bootstrap for testing hierarchical structures.
In Section \ref{sec-real}, we perform hypothesis testing for a linear attribute hierarchy in an educational assessment dataset and compare different testing procedures.
Finally Section \ref{sec-discuss} concludes with some discussions.

\section{Model Setup and Motivations}
\label{sec-model}

In this section, we introduce the model setup of CDMs and   the problem of interest. 
We start with some notations.
For an integer $K$, we use $[K]$ to denote the set $\{1,2,\dots,K\}$.
For two vectors $\bm{a} = (a_1,\dots,a_K)$ and $\bm{b}=(b_1,\dots,b_K)$ of the same length, we define a partial order ``$\succeq$" that
we say $\bm{a}\succeq\bm{b}$ if $a_k \geq b_k$ for $\forall \ k\in[K]$ and $\bm{a}\nsucceq \bm{b}$ otherwise. 
``$\preceq$" and ``$\npreceq$" are defined similarly.
We use $|\cdot|$ to denote the cardinality of a set.

\subsection{Cognitive Diagnosis Models}\label{chap1-sec-CDM}
We first introduce the model setup of CDMs and provide some illustrative examples. 
In CDMs, $J$ items are assumed to depend on $K$ latent attributes of interest.
In this work, both the responses and the latent attributes are assumed to be binary.
Based on a subject's observed responses to the items $\bm{R} = (R_1,\dots,R_J)\in\{0,1\}^J$, the latent attribute profile of the subject $\bm{\alpha} = (\alpha_1,\dots,\alpha_K)\in \{0,1\}^K$ denoting the profile of possession of the latent attributes needs to be inferred.
With $K$ latent binary attributes, there are $2^K$ possible latent attribute profiles, and we use $\bm{p}=(p_{\bm{\alpha}}:\bm{\alpha}\in\{0,1\}^K)$ to denote the proportion parameter vector for the latent attribute profiles which satisfies $p_{\bm{\alpha}}\in [0,1]$ and $\sum_{\bm{\alpha}\in\{0,1\}^K} p_{\bm{\alpha}}=1$. 
We use $\Delta^n$ to denote the standard $n$-simplex, that is, $\Delta^n = \{(t_0,\dots,t_n)\in \mathbb{R}^{n+1}:\sum_{i=0}^n t_i = 1, t_i\geq 0 \text{ for all }i\}$, 
and then we have $\bm{p}\in \Delta^{2^K-1}$.
In CDMs, the latent attribute profiles are assumed to follow a categorical distribution with proportion parameter vector $\bm{p}\in \Delta^{2^K-1}$.

A key ingredient of CDMs is a structural binary matrix, the so-called $Q$-matrix \citep{tatsuoka1990toward},  $\bm{Q}=(q_{j,k})\in\{0,1\}^{J\times K}$, which imposes constraints on items to reflect the dependence structure between the items and the latent attributes. 
Specifically, $q_{j,k}=1$ indicates that item $j$ requires (or depends on) attribute $k$. 
Then the $j$th row vector of $\bm{Q}$ denoted by $\bm{q}_j$ describes the full dependence of item $j$ on $K$ latent attributes.
Usually in applications, the $Q$-matrix is pre-specified by domain experts to reflect some scientific assumptions \citep{george2015cognitive, junker2001cognitive, von2005general}.
See Figure \ref{fig:Q} for illustration of the Q-matrix.

\begin{figure}[ht]
    \centering
    \subfigure{
    $\bm{Q} = \begin{pmatrix}
    	1 & 1 & 0 & 0\\
    	0 & 0 & 1 & 0 \\
    	0 & 1 & 0 & 1 \\
    	1 & 1 & 0 & 1 \\
    	0 & 0 & 1 & 1 \\
    \end{pmatrix}$}
    \hspace{1in}
    \subfigure{
    \begin{tikzpicture}[baseline= -13ex][scale=0.8]
    \path 
    node at (-2.6, -1.5) [place] (1) {$\alpha_1$}
    node at (-1.4, -1.5) [place] (2) {$\alpha_2$}
    node at (0, -1.5) [place] (3) {$\alpha_3$}
    node at (1.4, -1.5) [place] (4) {$\alpha_4$}
    node at (-3.5, -3) [place, fill={rgb:black,1;white,2},] (11) {$R_1$}
    node at (-2, -3) [place, fill={rgb:black,1;white,2},] (22) {$R_2$}
    node at (-0.5, -3) [place, fill={rgb:black,1;white,2},] (33) {$R_3$}
    node at (1, -3) [place, fill={rgb:black,1;white,2},] (44) {$R_4$}
    node at (2.5, -3) [place, fill={rgb:black,1;white,2},] (55) {$R_5$};
  \draw [->] (1) to (11);
  \draw [->] (1) to (44);
  \draw [->] (2) to (11);
  \draw [->] (2) to (33);
  \draw [->] (2) to (44);
  \draw [->] (3) to (22);
  \draw [->] (3) to (55);
  \draw [->] (4) to (33);
  \draw [->] (4) to (44);
  \draw [->] (4) to (55);
  \end{tikzpicture}
	}
    \caption{Illustration of Q-matrix}
    \label{fig:Q}
\end{figure}
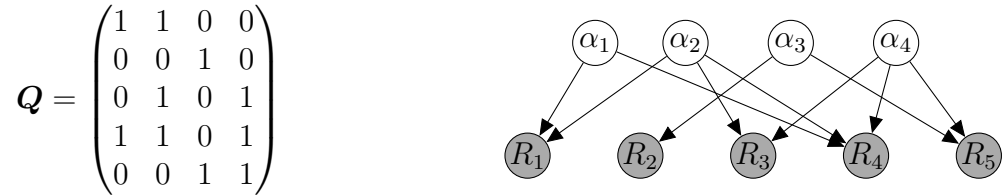{}

As in classical latent class analysis, given a subject's latent attribute profile $\bm{\alpha}$, the responses to $J$ items are assumed to be independent and follow Bernoulli distributions with parameters $\theta_{1,\bm{\alpha}},\dots,\theta_{J,\bm{\alpha}}$, which are called item parameters.
Specifically, $\theta_{j,\bm{\alpha}}:=\mathbb{P}(R_j=1\mid \bm{\alpha})$.
We use $\bm{\theta}_j = \big(\theta_{j,\bm{\alpha}}: \bm{\alpha}\in \{0,1\}^K\big)$ to denote the item parameter vector for the $j$th item, and $\bm{\Theta} = \big(\theta_{j,\bm{\alpha}}: j\in [J], \bm{\alpha}\in \{0,1\}^K \big)$ to denote the item parameter matrix.
Under the local independence assumption, the probability mass function of a subject's response vector $\bm{R} = (R_1,\dots,R_J)\in\{0,1\}^J$ then can be written as 
\begin{equation}
    \mathbb{P}(\bm{R}\mid \bm{\Theta},\bm{p}) = \sum_{\bm{\alpha}\in\{0,1\}^K} p_{\bm{\alpha}}\prod_{j=1}^J \theta_{j,\bm{\alpha}}^{R_j}(1-\theta_{j,\bm{\alpha}})^{1-R_j}.
\end{equation}

Under the CDM framework, for subject $i$, we have latent attribute profile $\bm{\alpha}_i$ to indicate subject $i$'s possession of $K$ attributes, and for item $j$, we have a structural vector $\bm{q}_j$ to reflect the item's dependence on $K$ latent attributes.
Under different model assumptions, the structural matrix $\bm{Q}$ puts constraints on item parameters in different ways.
One important common assumption in CDMs is that the item parameter $\theta_{j,\bm{\alpha}}$ only depends on whether the latent attribute profile $\bm{\alpha}$ contains the required attributes by item $j$, i.e., the attributes in the set $\mathcal{K}_j = \{k\in[K]:q_{j,k}=1\}$, which is called the \textit{set of required attributes of item j} in \cite{gu2019learning}. 
Therefore, for item $j$, the latent attribute profiles which are only different in the attributes outside of $\mathcal{K}_j$ would have the same item parameters. 
In such a way, the structural matrix $\bm{Q}$ forces some entries in the item parameter matrix $\bm{\theta}$ to be the same.
The dependencies of item parameters on the required attributed are modeled differently in different CDMs.
See Example \ref{ex-two} and Example \ref{ex-multi} for two popular families of such models.

\begin{example}[DINA and DINO Models]
\label{ex-two}
The DINA model \citep{junker2001cognitive} and the DINO model \citep{templin2006measurement} are two basic models in cognitive diagnosis.
The DINA model assumes a conjunctive ``AND" relationship among the latent binary attributes while the DINO model assumes a conjunctive ``OR" relationship.
We introduce a binary matrix $\bm{\Gamma} = \big(\Gamma_{j,\bm{\alpha}}:j\in [J], \bm{\alpha}\in\{0,1\}^K\big)\in\{0,1\}^{J\times 2^K}$, which is called the ideal response matrix.
In the DINA model, an ideal response is defined as $\Gamma^{\text{DINA}}_{j,\bm{\alpha}}=\mathbb{I}(\bm{\alpha}\succeq \bm{q}_j) = \prod_{k=1}^K \alpha_k^{q_{j,k}}$, indicating whether the latent profile contains all the required attributes.
In the DINO model, an ideal response is defined as $\Gamma^{\text{DINO}}_{j,\bm{\alpha}}=\mathbb{I}\big(\exists \ k \in [K], q_{j,k}=\alpha_k=1\big)$, indicating whether the latent profile contains any of the required attributes.
The item level uncertainties are characterized by two parameters, the slipping parameter and the guessing parameter.
For item $j$, the slipping parameter $s_j := \mathbb{P}(R_j = 0 \mid \Gamma_{j,\bm{\alpha}} = 1)$ denotes the probability of a subject giving a negative response despite possessing all the necessary skills; 
while the guessing parameter $g_j := \mathbb{P}(R_j = 1 \mid \Gamma_{j,\bm{\alpha}} = 0)$  denotes the probability of giving a positive response despite deficiency of some necessary skills.
Then the item parameter for item $j$ and latent profile $\bm{\alpha}$ can be written as 
$
\theta_{j, \bm{\alpha}} := \mathbb{P}(R_j = 1 \mid \bm{\alpha}) =  (1-s_j)^{\Gamma_{j,\bm{\alpha}}} g_j^{1-\Gamma_{j,\bm{\alpha}}}.
$

\end{example}

\begin{example}[GDINA model]
The Generalized DINA model  \citep[GDINA,][]{de2011generalized} is a more general model where all interactions among the required latent attributes are considered.
The item parameters for the GDINA model are written as
\begin{equation}
    \theta_{j,\bm{\alpha}}^{\text{GDINA}} = \beta_{j,0} + \sum_{k=1}^K \beta_{j,k} \alpha_k q_{j,k} + \sum_{k=1}^K \sum_{k' = k+1}^K \beta_{j,k,k'}\alpha_k \alpha_{k'}q_{j,k}q_{j,k'} +\cdots +\beta_{j,1,2,\dots,K} \prod_{k=1}^K \alpha_k q_{j,k}.
\end{equation}
The coefficients in the GDINA model can be interpreted as following: $\beta_{j,0}$ is the probability of a positive response for the most incapable subjects with no required attributes present; 
$\beta_{j,k}$ is the change in the probability of a positive response due to the main effect of $\alpha_k$;
$\beta_{j,k, k'}$ is the change in the probability of a positive response due to the interaction of $\alpha_k$ and $\alpha_k'$; 
$\beta_{j,1,2,\dots,K}$ is the change in the positive probability due to the interaction of all the latent attributes.
In the GDINA model, the intercept and main effects are typically assumed to be nonnegative, while the interactions can take negative values.
By incorporating all the interactions among the required attributes, the GDINA model is one of the most general CDMs.
\label{ex-multi}
\end{example}{}

\subsection{Problem and Motivations}
\label{sec-motivation}

\begin{figure}[ht] 
    \centering
    \subfigure{
    Linear
    }
    \hspace{2cm}
    \subfigure{ 
    Convergent
    }
    \hspace{2cm}
    \subfigure{ 
    Divergent
    }
    \hspace{2cm}
    \subfigure{ 
    Unstructured
    }
    \\
    \subfigure{
    \begin{tikzpicture}[scale=0.9]
    \path
    node at (0, -1.5) [place] (1) {$\alpha_1$}
    node at (0, -2.8) [place] (2) {$\alpha_2$}
    node at (0, -4.1) [place] (3) {$\alpha_3$}
    node at (0, -5.4) [place] (4) {$\alpha_4$};
    \draw [->, thick] (1) to (2);
    \draw [->, thick] (2) to (3);
    \draw [->, thick] (3) to (4);
    
    \path
    node at (5, -2) [place] (1) {$\alpha_1$}
    node at (4, -3.5) [place] (2) {$\alpha_2$}
    node at (6, -3.5) [place] (3) {$\alpha_3$}
    node at (5, -5.0) [place] (4) {$\alpha_4$};
    \draw [->, thick] (1) to (2);
    \draw [->, thick] (1) to (3);
    \draw [->, thick] (2) to (4);
    \draw [->, thick] (3) to (4);

    \path
    node at (10, -2) [place] (1) {$\alpha_1$}
    node at (9, -3.5) [place] (2) {$\alpha_2$}
    node at (11, -3.5) [place] (3) {$\alpha_3$}
    node at (11, -5.0) [place] (4) {$\alpha_4$};
    \draw [->, thick] (1) to (2);
    \draw [->, thick] (1) to (3);
    \draw [->, thick] (3) to (4);
    
    \path
    node at (15, -2) [place] (1) {$\alpha_1$}
    node at (14, -3.5) [place] (2) {$\alpha_2$}
    node at (15, -3.5) [place] (3) {$\alpha_3$}
    node at (16, -3.5) [place] (4) {$\alpha_4$};
    \draw [->, thick] (1) to (2);
    \draw [->, thick] (1) to (3);
    \draw [->, thick] (1) to (4);
    \end{tikzpicture}
    }
    \\
   	\hspace{-1cm}
    \subfigure{
    $
    \mathcal{A}_1 = 
    \begin{pmatrix}
    0 & 0 & 0 & 0\\
    1 & 0 & 0 & 0\\
    1 & 1 & 0 & 0\\
    1 & 1 & 1 & 0\\
    1 & 1 & 1 & 1\\ 
    \end{pmatrix}{}
    $
    }
    \hspace{0cm}
    \subfigure{
    $
    \mathcal{A}_2 = 
    \begin{pmatrix}
    0 & 0 & 0 & 0\\
    1 & 0 & 0 & 0\\
    1 & 1 & 0 & 0\\
    1 & 0 & 1 & 0\\
    1 & 1 & 1 & 0\\
    1 &1 & 1 &1\\ 
    \end{pmatrix}{}
    $
    }
    \hspace{0cm}
    \subfigure{
    $
    \mathcal{A}_3 = 
    \begin{pmatrix}
    0 & 0 & 0 & 0\\
    1 & 0 & 0 & 0\\
    1 & 1 & 0 & 0\\
    1 & 0 & 1 & 0\\
    1 & 0 & 1 & 1\\
    1 & 1 & 1 & 0\\
    1 & 1 & 1 & 1\\ 
    \end{pmatrix}{}
    $
    }
    \hspace{0cm}
    \subfigure{
    $
    \mathcal{A}_4 = 
    \begin{pmatrix}
    0 & 0 & 0 & 0\\
    1 & 0 & 0 & 0\\
    1 & 1 & 0 & 0\\
    1 & 0 & 1 & 0\\
    1 & 0 & 0 & 1\\
    1 & 1 & 1 & 0\\
    1 & 1 & 0 & 1\\
    1 & 0 & 1 & 1\\
    1 & 1 & 1 & 1\\ 
    \end{pmatrix}{}
    $
    }
    \\
    \subfigure{
    $|\mathcal{A}_1| = 5$
    }
    \hspace{2.3cm}
    \subfigure{
    $|\mathcal{A}_2| = 6$
    }
    \hspace{2.3cm}
    \subfigure{
    $|\mathcal{A}_3| = 7$
    }
    \hspace{2.3cm}
    \subfigure{
    $|\mathcal{A}_4| = 9$
    }

    \caption{Examples of hierarchical structures of latent attributes. For $i=1, \dots, 4$, each $\mathcal{A}_i$ represents the induced set of existent attribute profiles under the hierarchical structure above it, where each row in $\mathcal{A}_i$ represents an attribute profile $\bm{\alpha}$ with $p_{\bm{\alpha}}\neq 0$.}
    \label{fig:hier}
\end{figure}
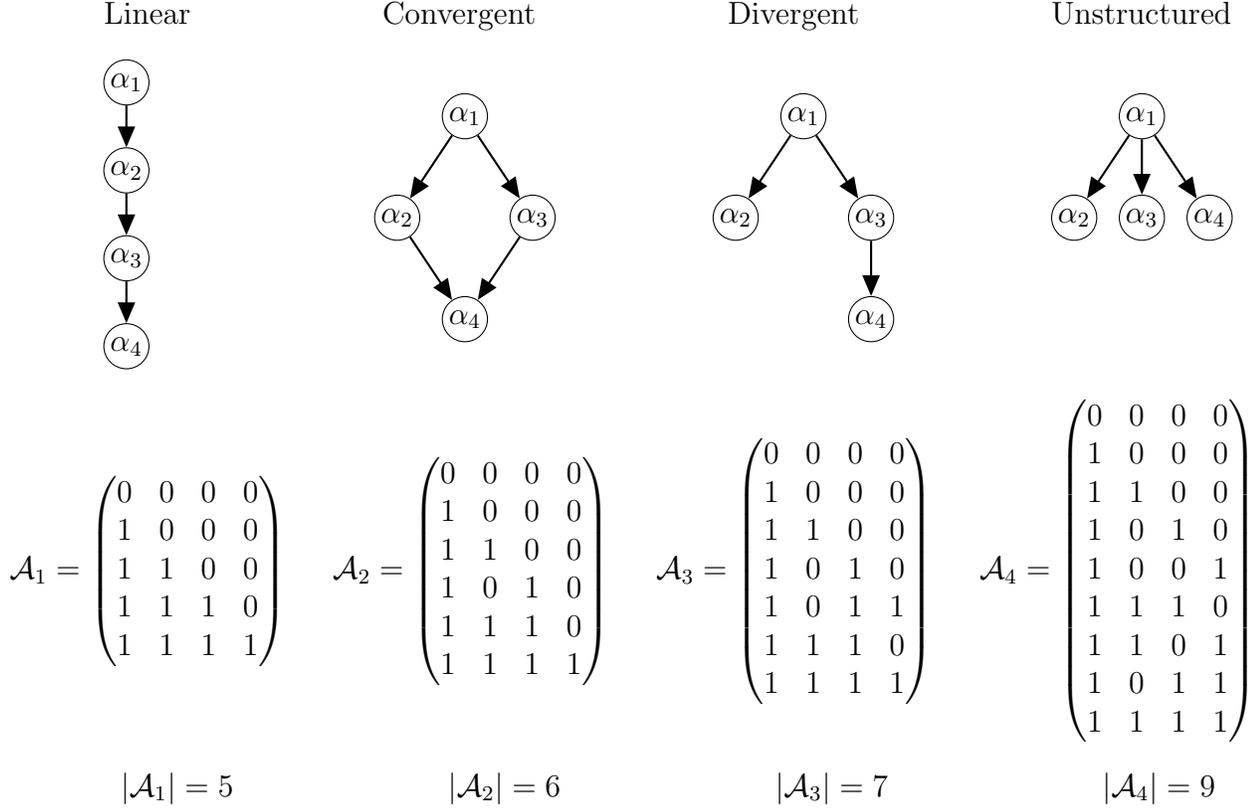

Researchers in many applications  assume  certain hierarchical structures among the latent attributes to characterize their dependence.
For example, in cognitive diagnosis modeling, the possession of lower-level skills is often regarded as the prerequisite for gaining higher-level skills \citep{leighton2004attribute, templin2014hierarchical}.
If some hierarchical structure exists, any latent profile $\bm{\alpha}$ that does not respect the hierarchy is deemed not to exist with the corresponding population proportion $p_{\bm{\alpha}}=0$. 
For $1\leq k \neq l \leq K$, we use $\alpha_k \xrightarrow{} \alpha_l$ (or $k\xrightarrow{}l$) to denote the hierarchy that attribute $\alpha_k$ is a prerequisite for attribute $\alpha_l$.
Under the hierarchy $\alpha_k \xrightarrow{} \alpha_l$, the latent profiles with $\alpha_l=1$ but $\alpha_k=0$ will not exist in the population and therefore we have $p_{\bm{\alpha}}=0$ if $\alpha_l=1$ but $\alpha_k=0$. 
The set of prerequisite relationships is denoted by $\mathcal{E} = \{k\xrightarrow{}l: \text{ attribute }k \text{ is a prerequisite for } l, \ 1\leq k \neq l \leq K\}$, and the induced set of existing latent attribute profiles is denoted by $\mathcal{A}=\{\bm{\alpha}\in\{0,1\}^K:p_{\bm{\alpha}}\neq 0 \text{ under }\mathcal{E}\}$.
It is noted that an attribute hierarchy results in the sparsity of the proportion parameter vector, which will reduce the number of model parameters especially when $K$ is large.
Example hierarchical structures and the corresponding induced latent profile sets are shown in Figure \ref{fig:hier}.

In many applications, certain hierarchical structures are postulated by domain experts.
In this paper, we consider the problem of hypothesis testing for the existence of such pre-specified latent hierarchical structures.
As we illustrated above, the hierarchical structure of the latent attributes results in the sparsity structure of the proportion parameter vector for the latent attribute profiles, since the latent profiles that do not follow the hierarchical structure will not exist in the population.
Therefore the problem of testing latent hierarchy is equivalent to testing the sparsity structure of the proportion parameter vector.
More formally, we aim to test the following hypothesis:
\[
\text{H}_0: p_{\bm{\alpha}}=0, \ \forall \bm{\alpha}\notin \mathcal{A}_0 \text{ under hierarchy } \mathcal{E}_0,
\]
where $\mathcal{E}_0$ is the hierarchical structure under the null hypothesis and $\mathcal{A}_0$ is the induced latent attribute profile set under $\mathcal{E}_0$.

Even though CDMs can be viewed as a special family of finite mixture models, 
there is a key difference between testing hierarchical structures in CDMs and testing the number of components in finite mixture models. 
When testing the number of components in finite mixture models, there is no restrictions on the components' distributions.
See \cite{chen2017finite} for a review of testing the number of components in finite mixture models.
However, when testing latent hierarchical structures, we are in fact testing whether the proportion parameters of the nonexistent latent attribute profiles corresponding to the hierarchy are zeros.
Moreover, the constraints imposed by the structural $Q$-matrix makes it more restrictive and complicated.

\section{Testability Requirements and Conditions}
\label{sec-test}

Before we introduce concrete testing procedures,
we first need to understand when the hierarchical structures are testable.
For instance, consider the case when the item parameters are the same for two latent attribute profiles and we want to test the nonexistence of one of them.
In this situation, we cannot distinguish these two profiles and thus cannot identify their proportion parameters, 
not mention to test whether the corresponding proportion is zero.
Example \ref{ex:ex1} provides an illustrative example.
Therefore, the testability issue is of fundamental importance before performing concrete testing procedures.
Moreover, the testability conditions would also provide guidance for practitioners and scientific researchers to design experiments.

\begin{example}
Assume that there are two latent attributes of interest and there is a linear attribute hierarchy $\mathcal{E}_0 =\{1\xrightarrow{}2\}$, which results in the induced attribute profile set $\mathcal{A}_0=\{(0,0),(1,0),(1,1)\}$. If the $Q$-matrix is specified as,
\[
\bm{Q} = 
\begin{pmatrix}
1 & 0 \\
1 & 1 \\
1 & 0 \\
1 & 1 \\
1 & 1 \\
\end{pmatrix},
\]

then under the DINA model assumption, the item parameter vector for the latent profile $(0,0)$, $\bm{\theta}_{(0, 0)}$, would be the same as $\bm{\theta}_{(0, 1)}$.
In this case, we cannot distinguish the profiles $(0,0)$ and $(0,1)$, 
and therefore the proportion parameters $p_{(0, 0)}$ and $p_{(0, 1)}$ cannot be identified. 
Furthermore, the induced profile set $\mathcal{A}_0$ is not identifiable, which makes the latent hierarchical structure untestable.
\label{ex:ex1}
\end{example}

To ensure the testability of hierarchical structures, some conditions need to be met. 
Before we dive into these conditions,
let's first define the concept of the testability of latent hierarchies.

\begin{definition}[strict testability of $\mathcal{E}_0$]
Given the $Q$-matrix and certain cognitive diagnosis model assumptions, consider the following hypothesis testing:
\begin{center}
$\text{H}_0$: the latent attributes respect the hierarchy  $\mathcal{E}_0$,

$\text{H}_1$: the latent attributes do not respect the hierarchy $\mathcal{E}_0$.    
\end{center}{}

\noindent Then the latent hierarchy $\mathcal{E}_0$ is said to be testable, if there is no parameter under the alternative hypothesis gives the same distribution as the parameters under the null hypothesis.
\end{definition}{}

In fact, the testability is closely related to the identifiability of CDMs \citep{xu2016identifiability,xu2017identifiability}.
The identifiability refers to that if two parameters give the same distribution, then the two parameters must be the same.
Nevertheless, the testability of hierarchical structure is actually less restrictive compared with the identifiability. 
In testing latent hierarchies, we only need to distinguish the latent attribute profiles under the null hierarchical structure with the others under the alternative, 
while in terms of the identifiability, we need to identify all the model parameters and all the latent attribute profiles. 
Therefore the concept of testability is weaker than the definitions of identifiability. 
In particular, identifiability is a sufficient but not necessary condition for testability. 

We first consider the DINA model.
For the DINA model, since the item parameters only depend on the highest interactions among the required latent attributes, we have equivalent $Q$-matrices under hierarchical structures. 
Here we say two $Q$-matrices are equivalent under hierarchical structure $\mathcal{E}$, denoted by $\mathbf{Q}_1 \overset{\mathcal{E}}{\sim} \mathbf{Q}_2$, if they give the same item parameter matrices, that is, $\boldsymbol{\Theta}(\mathbf{Q}_1,\mathcal{A}_{\mathcal{E}}) = \boldsymbol{\Theta}(\mathbf{Q}_2,\mathcal{A}_{\mathcal{E}})$, where $\mathcal{A}_{\mathcal{E}}$ is the induced latent attribute profile set under hierarchy $\mathcal{E}$.
For example, consider three latent attributes with a linear hierarchy, that is, $\mathcal{E} = \big\{1\rightarrow 2 \rightarrow 3 \big\}$.
We have 
\begin{equation}
	\mathbf{Q}^{(1)} = \begin{pmatrix}
		1 & 0 & 0 \\
		0 & 1 & 0 \\
		0 & 0 & 1 \\
	\end{pmatrix} \overset{\mathcal{E}}{\sim}
	\mathbf{Q}^{(2)} = \begin{pmatrix}
		1 & 0 & 0 \\
		1 & 1 & 0 \\
		1 & 1 & 1 \\
	\end{pmatrix} \overset{\mathcal{E}}{\sim}
	\mathbf{Q}^{(*)} = \begin{pmatrix}
		1 & 0 & 0 \\
		* & 1 & 0 \\
		* & * & 1 \\
	\end{pmatrix}, 
	\label{fig:equivalent-Q}
\end{equation}
where ``$*$" can be either 0 or 1.
Based on this observation, following \cite{gu2021identifiability}, we introduce two useful operations on the $Q$-matrix.

\begin{definition}
Given an attribute hierarchy $\mathcal{E}$ and a $Q$-matrix $\bm{Q}$. 
For any $q_{j,l} = 1$ and $k\rightarrow l$, set $q_{j,k}$ to 0 and obtain a modified matrix $\mathcal{S}^{\mathcal{E}}(\bm{Q})$.
This $\mathcal{S}^{\mathcal{E}}(\bm{Q})$ is called the ``sparsified" version of $\bm{Q}$.
\label{def-densify}	
\end{definition}

\begin{definition}
Given an attribute hierarchy $\mathcal{E}$ and a $Q$-matrix $\bm{Q}$. 
For any $q_{j,l} = 1$ and $k\rightarrow l$, set $q_{j,k}$ to 1 and obtain a modified matrix $\mathcal{D}^{\mathcal{E}}(\bm{Q})$.
This $\mathcal{D}^{\mathcal{E}}(\bm{Q})$ is called the ``densified" version of $\bm{Q}$.
\label{def-sparsify}
\end{definition}

As we discussed previously, the identifiability conditions are sufficient conditions for the testability.
We present some identifiability results in \cite{gu2021identifiability} for the DINA model.

\begin{proposition}[strict testability for the DINA model]
Consider a DINA model with a given $\bm{Q}$. 
A hierarchy $\mathcal{E}_0$ is testable if $
\bm{Q}$ satisfies the following conditions:
\renewcommand\labelenumi{(\theenumi)}
\begin{enumerate}
    \item 
    $\bm{Q}$ contains a $K\times K$ identity submatrix $I_K$.
   	(Without loss of generality, assume the first $K$ rows of $\bm{Q}$ form $I_K$ and denote the remaining submatrix of $\bm{Q}$ by $\bm{Q}^*$.)
    \item 
    $\mathcal{S}^{\mathcal{E}_0}(\bm{Q})$, the sparsified version of $\bm{Q}$, has at least three entries of ``1" in each column.
    \item 
    $\mathcal{D}^{\mathcal{E}_0}(\bm{Q}^*)$, the densified version of $\bm{Q}^*$,  contains $K$ distinct column vectors.
\end{enumerate}{}
\label{prop-dina}
\end{proposition}

Among the above conditions, condition (1) is in fact necessary to ensure the testability of any hierarchy $\mathcal{E}$, which is however not satisfied in Example \ref{ex:ex1}.
Since the $Q$-matrix in Example \ref{ex:ex1} does not contain $(0, 1)$, we can not distinguish the latent attribute profiles $(0,1)$ and $(0,0)$, making the hierarchy not testable.
We revisit Example \ref{ex:ex1} with a different $Q$-matrix and demonstrate its testability.

\begin{example}[Example \ref{ex:ex1} revisited]
Consider the same setting as in Example \ref{ex:ex1}, but with a different Q-matrix specified as below:
\[
\bm{Q} = \begin{pmatrix}
	\bm{I}_2\\
	\bm{Q}^*
 \end{pmatrix}, \mbox{ where } 
  \bm{I}_2 = \begin{pmatrix}
 1 & 0\\
 0 & 1
 \end{pmatrix}    \mbox{and  }
   \bm{Q}^* = \begin{pmatrix}
	1 & 0 \\
	0 & 1 \\
	1 & 0 \\
	1 & 1
 \end{pmatrix}.
\]
Then the modified Q matrices are:
\[
\mathcal{S}^{\mathcal{E}_0}(\bm{Q}) = \begin{pmatrix}
 1 & 0 \\
 0 & 1 \\
 1 & 0 \\
 0 & 1 \\
 1 & 0 \\
 \textcolor{blue}{\bm{0}} & 1\\
 \end{pmatrix},
 \quad
\mathcal{D}^{\mathcal{E}_0}(\bm{Q}^*) =  \begin{pmatrix}

 1 & 0 \\
 \textcolor{blue}{\bm{1}} & 1 \\
 1 & 0 \\
 1 & 1 \\
 \end{pmatrix}.
\]
Since the $Q$-matrix contains an identity matrix $\bm{I}_2$, the sparsified version $\mathcal{S}^{\mathcal{E}_0}(\bm{Q})$ has three ``1" entries in each column, and the densified version $\mathcal{D}^{\mathcal{E}_0} (\bm{Q}^*)$ contains two distinct columns, 
all three conditions in Theorem 1 are satisfied. 
Therefore, for a DINA model with this $Q$-matrix, the linear hierarchy $\mathcal{E}_0$ is strictly testable.
\label{ex-dina-2}
\end{example}

Given an attribute hierarchy $\mathcal{E}_0$, 
if we are interested in testing a subset of prerequisite relations conditioned on that the other prerequisite relations are assumed, we can further relax Condition (1) in Proposition \ref{prop-dina}.
For example, assume that there are three latent attributes and the full hierarchical structure is $\mathcal{E}_0 = \{1\rightarrow 2 \rightarrow \ 3\} $.
If we are interested in testing $\mathcal{E} = \{1 \rightarrow 2\}$ given $\mathcal{E}_0 \setminus \mathcal{E} = \{2 \rightarrow 3\}$, that is $\text{H}_0: \ \mathcal{E}_0$ vs. $\text{H}_1: \ \mathcal{E}_0 \setminus \mathcal{E}$, we can relax Condition (1) in Proposition \ref{prop-dina} to Condition (1*) in Corollary \ref{corollary-dina-2}.

\begin{corollary}
Consider a DINA model with a given $\bm{Q}$ and a given attribute hierarchy $\mathcal{E}_0$.
Suppose we are interested in testing a subset $\mathcal{E} \subset \mathcal{E}_0$ given that $\mathcal{E}_0 \setminus \mathcal{E}$ has already been assumed.
Then it is testable if Condition (2) and (3) in Proposition \ref{prop-dina} and the following condition are satisfied:
\begin{enumerate}
	\item[($1^*$)]
	$\mathcal{S}^{\mathcal{E}_0}(\bm{Q})$, the sparsified version of $\bm{Q}$, contains an identity submatrix $I_K$ and 
	for any attribute $\alpha_k$ involved in $\mathcal{E}$, there is an item which is only targeted on this attribute.
\end{enumerate}
\label{corollary-dina-2}

\end{corollary}

The proof of Corollary \ref{corollary-dina-2} directly follows Theorem 1 in \cite{gu2021identifiability}.
As we mentioned previously, the testability is a weaker requirement than the identifiability, in that we only need to differentiate the latent attribute profiles between the null and alternative hypothesis for testability.
We next provide an example in which the hierarchical structure is testable but the model is not identifiable.

\begin{example}[Testability vs. Identifiability]
\label{ex-test-identify}
Consider a DINA model with three latent attributes.
Further assume the slipping and guessing parameters are known.
We want to test the linear hierarchy which is specified as $\mathcal{E}_0 = \{1 \rightarrow 2 \rightarrow 3 \}$.
Then the induced latent attribute profile set is $\mathcal{A}_0 = \{(0,0,0), (1,0,0), (1,1,0), (1,1,1)\}$.
Consider the $Q$-matrix:
\[
\bm{Q} = \begin{pmatrix}
	0 & 1 & 0\\
	0 & 0 & 1\\
	1 & 1 & 0\\
	1 & 1 & 1
\end{pmatrix}.
\]	
We denote the set of latent attribute profiles that do not exist under the hierarchy as $\mathcal{A}_0^c = \{(0,1,0), (0,0,1), (1,0,1), (0,1,1)\}$.
Then under the specified $Q$-matrix, the ideal response matrices for $\mathcal{A}_0$ and $\mathcal{A}_0^c$ are:
\begin{equation*}
	\bm{\Gamma}^{\mathcal{A}_0} = \begin{pmatrix}
		0 & 0 & 1 & 1 \\
		0 & 0 & 0 & 1 \\
		0 & 0 & 1 & 1 \\
		0 & 0 & 0 & 1 \\
	\end{pmatrix},
	\quad 
	\bm{\Gamma}^{\mathcal{A}_0^c} = \begin{pmatrix}
		1 & 0 & 0 & 1 \\
		0 & 1 & 1 & 1 \\
		0 & 0 & 0 & 0 \\
		0 & 0 & 0 & 0 \\
	\end{pmatrix}.
\end{equation*}
Therefore the ideal response vectors for classes in $\mathcal{A}_0$ and $\mathcal{A}_0^c$ are different, making the hierarchical structure testable.
However, the ideal response vectors for $(0,0,0)$ and $(1,0,0)$ are the same, and those for $(0,0,1)$ and $(1,0,1)$ are also the same, making the model not identifiable.	
\end{example}

\begin{example}[Condition on Alternative]
Consider a DINA model with three latent attributes.
The hierarchical structure is specified as $\mathcal{E}_0 = \{1 \rightarrow 2 \rightarrow 3 \}$.
Then the induced latent attribute profile set is $\mathcal{A}_0 = \{(0,0,0), (1,0,0), (1,1,0), (1,1,1)\}$.
We are interested in testing $\{1\rightarrow 2\}$ given $\{2 \rightarrow 3\}$.
Consider the $Q$-matrix:
\[
\bm{Q} = \begin{pmatrix}
	1 & 0 & 0\\
	0 & 1 & 0\\
	1 & 0 & 0\\
	0 & 1 & 0\\
	1 & 0 & 0\\
	1 & 1 & 0\\
	1 & 0 & 1 \\
	0 & 1 & 1\\
	1 & 1 & 1
\end{pmatrix}.
\]	
Since $\bm{\theta}_{(0,0,0)} = \bm{\theta}_{(0,0,1)}$, we cannot distinguish the latent profiles $(0,0,0)$ and $(0,0,1)$, and thus not all the latent profiles are identifiable. 
However, the conditions in Proposition \ref{prop-dina} are satisfied, so the hierarchical structure $\{1\rightarrow 2\}$ given $\{2\rightarrow 3\}$ is testable.
	
\end{example}

For more general CDMs, we adapt the identifiability results from \cite{gu2019learning} to establish sufficient conditions for the testability of latent attribute hierarchies.
Following the same notations in \cite{gu2019learning},
we first introduce the so-called constraint matrix $\bm{\Gamma}$.
The constraint matrix for a set of latent attribute profiles  $\mathcal{A}$ is defined as $\bm{\Gamma}^{\mathcal{A}} = \Big(\mathbb{I}(\bm{\alpha}\succeq \bm{q}_j): \bm{\alpha}\in \mathcal{A}, j\in[J] \Big) \in \{0, 1\}^{J\times |\mathcal{A}|}$, which is a binary matrix indicating whether an attribute profile $\bm{\alpha}\in\mathcal{A}$ possesses all the required attributes of item $j$.
Note that for the DINA model, the constraint matrix is also its ideal response matrix, while they are not the same for the DINO model. The defined constraint matrix is used  as a  tool to study the testability conditions for general CDMs.
Based on the constraint matrix, we define a partial order among the latent attribute profiles ``$\succeq_{S}$" for any subset of items $S\subset [J]$. 
For $\bm{\alpha}, \bm{\alpha}' \in \mathcal{A}$, we say $\bm{\alpha}\succeq_{S}\bm{\alpha}'$ under $\bm{\Gamma}^{\mathcal{A}}$ if $\Gamma_{j,\bm{\alpha}}^{\mathcal{A}} \geq \Gamma_{j,\bm{\alpha}'}^{\mathcal{A}}$ for $j \in S$. 
And for two item sets $S_1$ and $S_2$, we say ``$\succeq_{S_1}=\succeq_{S_2}$" if for any $\bm{\alpha}, \bm{\alpha}' \in \mathcal{A}$, we have $\bm{\alpha}\succeq_{S_1}\bm{\alpha}'$ if and only if  $\bm{\alpha}\succeq_{S_2}\bm{\alpha}'$.
Example \ref{ex-partial-order} provides illustrations for the partial orders.

\begin{example}
 	\label{ex-partial-order}
	Consider two latent attributes with a linear hierarchical structure and the following $Q$-matrix:
	\[
		\bm{Q} = \begin{pmatrix}
			1 & 0\\
			0 & 1\\
			1 & 0\\
			1 & 1\\
		\end{pmatrix}.
	\]	
	The corresponding constraint matrix for the latent attribute profile set $\mathcal{A} = \{(0,0), (1, 0), (1, 1)\}$ is:
	\[
		\bm{\Gamma}^{\mathcal{A}} = \begin{blockarray}{ccc}
    		(0,0) & (1,0) & (1, 1)\\
    		\begin{block}{(ccc)}
       			0 & 1 & 1\\
       			0 & 0 & 1\\
       			0 & 1 & 1\\
       			0 & 0 & 1\\
    		\end{block}
    	\end{blockarray}\ .
    \]  
    For the item set $S = \{1, 2\}$, we can see that $\Gamma_{j,(1,0)} \geq \Gamma_{j,(0,0)}$ and $\Gamma_{j,(1,1)} \geq \Gamma_{j,(1,0)}$ for $j \in S$. 
    Therefore $(1, 0) \succeq_{S} (0, 0)$ and $(1, 1) \succeq_{S} (1, 0)$.
    Moreover, if we take $S_1 = \{1, 2\}$ and $S_2 = \{3, 4\}$, then for any $\bm{\alpha}, \bm{\alpha}' \in \mathcal{A}$, we have $\bm{\alpha}\succeq_{S_1}\bm{\alpha}'$ if and only if  $\bm{\alpha}\succeq_{S_2}\bm{\alpha}'$.
	Therefore, $\succeq_{S_1}=\succeq_{S_2}$.
\end{example}

In the following testability results for general CDMs, we focus on equal size cases or under-fitted cases when $|\mathcal{A}| \leq |\mathcal{A}_0|$, where $\mathcal{A}_0$ is the set of the latent attribute profiles under the null hypothesis and $\mathcal{A}$ is the set of the latent attribute profiles under the alternative hypothesis. 
Note that for overfitted cases with $|\mathcal{A}| > |\mathcal{A}_0|$, if $\mathcal{A}$ and $\mathcal{A}_0$ lead to the same distribution, the model complexity of $\mathcal{A}$ is larger than that of $\mathcal{A}_0$,
and therefore practically we can still distinguish them 
using information-based criteria or penalized likelihood methods.

\begin{proposition}[strict testability for general CDMs]
Consider a general CDM with a given $\bm{Q}$ and an arbitrary hierarchy $\mathcal{E}_0$.
The hierarchy is testable when the alternative is restricted to the latent profile sets of the same or a smaller size than that under the null hypothesis, 
if the following conditions of the constraint matrix $\bm{\Gamma}^{\mathcal{A}_0}$ corresponding to the induced latent profile set $\mathcal{A}_0$ under the hierarchy $\mathcal{E}_0$ are satisfied:

\renewcommand\labelenumi{(\theenumi)}
\begin{enumerate}
    \item There exist two disjoint item sets $S_1$ and $S_2$, such that $\bm{\Gamma}^{(S_i,\mathcal{A}_0)}$ has distinct column vectors for $i=1,2$ and ``$\succeq_{S_1}=\succeq_{S_2}$" under $\bm{\Gamma}^{\mathcal{A}_0}$.
    \item For any $\bm{\alpha},\ \bm{\alpha}'\in \mathcal{A}_0$ where $\bm{\alpha}'\succeq_{S_i}\bm{\alpha}$ under $\bm{\Gamma}^{\mathcal{A}_0}$ for $i=1$ or 2, there exists some $j\in \big(S_1\cup S_2\big)^c$ such that $\Gamma_{j,\bm{\alpha}}^{\mathcal{A}_0}\neq \Gamma_{j,\bm{\alpha}'}^{\mathcal{A}_0}$.
    \item Any column vector of $\bm{\Gamma}^{\mathcal{A}_0}$ is different from any column vector of $\bm{\Gamma}^{\mathcal{A}_0^c}$, where $\mathcal{A}_0^c = \{0,1\}^K\setminus \mathcal{A}_0$
\end{enumerate}{}
\label{prop-gdina}
\end{proposition}

Based on the conditions in Proposition \ref{prop-gdina}, one can see that having three identity submatrices in the $Q$-matrix is sufficient for testability.
However, having several identity submatrices is in fact a strong requirement in practice.
Under a general CDM, these conditions can be further relaxed if we consider $\mathcal{E}_0$ to be testable with the true model parameter ranging almost everywhere in the restricted parameter space except a set of Lebesgue measure zero. 
Specifically, we have the following definition of generic testability.

\begin{definition}[generic testability of $\mathcal{E}_0$]
Denote the parameter space under $\mathcal{E}_0$ by $\bm{\Omega}_0$.
The latent hierarchy $\mathcal{E}_0$ is said to be generically testable, if there exists a subset $\bm{\mathcal{V}}$ of $\bm{\Omega}_0$ that has Lebesgue measure zero, such that there is no parameter under the alternative hypothesis gives the same distribution as the parameters in $\bm{\Omega}_0 \setminus \bm{\mathcal{V}}$.

\end{definition}

For generic testability, following the generic identifiability results in   \cite{gu2019learning} and \cite{gu2020partial},
 a nice corollary can be derived where the requirements are directly characterized by the structure of the $Q$-matrix.

\begin{corollary}
If the $Q$-matrix satisfies the following conditions, then for any hierarchy $\mathcal{E}_0$ such that the induced latent attribute profile set $\mathcal{A}_0$ satisfies Condition (3) in Proposition \ref{prop-gdina}, the hierarchy $\mathcal{E}_0$ is generically testable:
\renewcommand\labelenumi{(\theenumi)}
\begin{enumerate}
\item The $Q$-matrix contains two $K\times K$ sub-matrices $\bm{Q}_1$ and $\bm{Q}_2$, such that for $i = 1, 2$,
\[
\bm{Q} = \begin{pmatrix}
\bm{Q}_1\\
\bm{Q}_2\\
\bm{Q}'
 \end{pmatrix}_{J\times K};
 \quad 
\bm{Q}_i = \begin{pmatrix}
 1 & * & \cdots & *\\
 * & 1 & \cdots & *\\
 \vdots & \vdots & \ddots & \vdots \\
 * & * & \cdots & 1 
 \end{pmatrix}_{K\times K}, i = 1,2,
\]
where each ``$*$'' can be either zero or one.
\item With $\bm{Q}$ in the form as above, $\sum_{j=2K+1}^J q_{j,k}\geq 1$ for each $k\in [K]$. 
\end{enumerate}

\label{corollary-gdina-gene}
\end{corollary}

By relaxing strict testability to generic testability, less stringent conditions in Corollary \ref{corollary-gdina-gene} have been established.
Moreover, the requirements in Corollary \ref{corollary-gdina-gene} can be checked directly from the $Q$-matrix, making it easier to use in practice.
Next, we present an illustrative example about strict testability and generic testability of general CDMs.

\begin{example}
Consider a general CDM setting with two latent attributes and a linear attribute hierarchy $\mathcal{E}_0 = \{1 \rightarrow 2\}$. Consider the $Q$-matrix:
\[
\bm{Q} = \begin{pmatrix}
	\bm{I}_2\\
	\bm{I}_2\\
	\bm{Q}'\\
\end{pmatrix};
\quad 
\bm{Q}' = \begin{pmatrix}
	1 & 0\\
	1 & 1
\end{pmatrix}.
\]
By directly looking at the $Q$-matrix, we know the conditions in Corollary \ref{prop-gdina} are satisfied and therefore the hierarchical structure is generically testable.
Moreover, the constraint matrix under attribute hierarchy $\mathcal{E}_0 = \{1\rightarrow 2\}$ is 
\[
\bm{\Gamma}^{\mathcal{A}_0} =     \begin{blockarray}{ccc}
    (0,0) & (1,0) & (1,1) & \\
    \begin{block}{(ccc)}
       0 & 1 & 1  \\
       0 & 0 & 1  \\
       0 & 1 & 1  \\
       0 & 0 & 1  \\
       0 & 1 & 1  \\
       0 & 0 & 1  \\
    \end{block}
    \end{blockarray}\ ;
    \quad 
    \bm{\Gamma}^{\mathcal{A}_0^c} = \begin{blockarray}{c}
    (0,1)\\
    \begin{block}{(c)}
       0  \\
       1  \\
       0  \\
       1  \\
       0  \\
       0  \\
    \end{block}
	\end{blockarray}.
\]
If we set $S_1 = \{1,2\},\ S_2 = \{3,4\}$, then $\bm{\Gamma}^{(S_i, \mathcal{A}_0)}$ has distinct columns for $i = 1, 2$.
Moreover, ``$\succeq_{S_1}=\succeq_{S_2}$" under $\bm{\Gamma}^{\mathcal{A}_0}$.
For $(1, 0) \succeq_{S_i} (0, 0)$ for $i = 1$ or 2, we have $\Gamma_{5,(1, 0)}^{\mathcal{A}_0} \neq \Gamma_{5, (0, 0)}^{\mathcal{A}_0}$.
For $(1, 1) \succeq_{S_i} (1, 0)$ for $i = 1$ or 2, we have $\Gamma_{6 ,(1, 1)}^{\mathcal{A}_0} \neq \Gamma_{6, (1, 0)}^{\mathcal{A}_0}$.
For $(1, 1) \succeq_{S_i} (0, 0)$ for $i = 1$ or 2, we have $\Gamma_{6 ,(1, 1)}^{\mathcal{A}_0} \neq \Gamma_{6, (0, 0)}^{\mathcal{A}_0}$.
Finally, the columns of $\bm{\Gamma}^{\mathcal{A}_0}$ are different from that of $\bm{\Gamma}^{\mathcal{A}_0^c}$.
Therefore, based on the constraint matrix, we can see that the conditions in Proposition \ref{prop-gdina} are met, and thus the linear attribute hierarchy is also strictly testable. 
	\label{ex-dina-3}
\end{example}

\begin{example}
Consider a general CDM setting with three latent attributes and a linear hierarchical structure $\mathcal{E}_0 = \{1 \rightarrow 2 \rightarrow 3 \}$.
The induced latent attribute profile set under $\mathcal{E}_0$ is $\mathcal{A}_0 = \{(0,0,0), (1,0,0), (1,1,0), (1,1,1)\}$.
We denote the complement set of latent attribute profiles as $\mathcal{A}_0^c = \{(0,1,0), (0,0,1), (1,0,1), (0,1,1)\}$.
Consider the $Q$-matrix:
\[
\bm{Q} = \begin{pmatrix}
	1 & 1 & 0\\
	0 & 1 & 0\\
	0 & 0 & 1\\
	1 & 1 & 0\\
	0 & 1 & 0\\
	0 & 0 & 1\\
	1 & 1 & 0\\
	1 & 0 & 1\\
	1 & 1 & 1\\
\end{pmatrix}.
\]	
Under the specified $Q$-matrix, the constraint matrices for $\mathcal{A}_0$ and $\mathcal{A}_0^c$ are:
\begin{equation*}
	\bm{\Gamma}^{\mathcal{A}_0} = \begin{pmatrix}
		0 & 0 & 1 & 1 \\
		0 & 0 & 1 & 1 \\
		0 & 0 & 0 & 1 \\
		0 & 0 & 1 & 1 \\
		0 & 0 & 1 & 1 \\
		0 & 0 & 0 & 1 \\
		0 & 0 & 1 & 1 \\
		0 & 0 & 0 & 1 \\
		0 & 0 & 0 & 1 
	\end{pmatrix},
	\quad 
	\bm{\Gamma}^{\mathcal{A}_0^c} = \begin{pmatrix}
		0 & 0 & 0 & 0 \\
		1 & 0 & 0 & 1 \\
		0 & 1 & 1 & 1 \\
		0 & 0 & 0 & 0 \\
		1 & 0 & 0 & 1 \\
		0 & 1 & 1 & 1 \\
		0 & 0 & 0 & 0 \\
		0 & 0 & 1 & 0 \\
		0 & 0 & 0 & 0
	\end{pmatrix}.
\end{equation*}
Based on the specified $Q$-matrix and the corresponding constraint matrices, one can easily see that the conditions in Corollary \ref{corollary-gdina-gene} are satisfied and therefore the hierarchical structure is generically testable.
However, Condition (1) in Proposition \ref{prop-gdina} is not satisfied and the model is not strictly identifiable  since $\bm{\Gamma}^{(0,0,0)}$  and $\bm{\Gamma}^{(1,0,0)}$ are the same.
\end{example}

\section{Likelihood Ratio Test}
\label{sec-LRT}

With the sufficient conditions for the testability of the hierarchical structures specified in Section \ref{sec-test},  the next question becomes how to conduct the hypothesis testing.
As we illustrated in Section \ref{sec-motivation}, when some hierarchical structure exists, the number of truly existing latent attribute profiles will be less than $2^K$, and the corresponding model will be a nested model of the full model with all possible latent attribute profiles.
Testing the latent hierarchical structure is then equivalent to testing the sparsity structure of the proportion parameter vector.
A popular choice of testing a nested model is the likelihood ratio test with an asymptotic Chi-squared distribution under some regularity conditions.
One commonly assumed regularity condition is that the true parameter vector is in the interior of the parameter space.
However, in our testing problem, the true proportion parameter vector $\bm{p}$ lies on the boundary of the simplex under the null hypothesis, making the conventional Chi-squared limiting distribution no longer hold.
In this section, we review the nonstandard asymptotic behaviors of the LRT statistic and provide statistical insights on the failures of such limiting distributions under practical conditions.
Then we propose to use resampling-based methods to test hierarchical structures and  conduct a comprehensive simulation study to compare different testing procedures.

\subsection{Failure of Limiting Distribution of LRT}
\label{sec-limiting}

When the parameter of the null model lies on the boundary of the parameter space, the LRT statistic has been shown to often follow a mixture of $\chi^2$ distributions asymptotically \citep{self1987asymptotic}.
We first present some general asymptotic theories on the LRT statistic under such nonstandard conditions and discuss the application to our testing problem for latent hierarchical structures.

Let $f(\bm{x};\bm{\theta})$ be the probability density function of a random variable $\bm{X}$, 
where $\bm{\theta} = (\theta_1, ..., \theta_p)$ takes values in the parameter space $\bm{\Omega}$, a subset of $\mathbb{R}^p$. 
When the model is identifiable, distinct values of $\bm{\theta}$ correspond to distinct probability distributions.
Let $\bm{x}_1, ..., \bm{x}_N$ be $N$ independent observations of $\bm{X}$ and denote the log-likelihood function, $\sum_{i=1}^N \log[f(\bm{x}_i;\bm{\theta})]$, by $l_N(\bm{\theta})$.
Consider the hypothesis testing
\[
\text{H}_0: \bm{\theta}_0 \in \bm{\Omega}_0 \quad vs \quad \text{H}_1: \bm{\theta}_0 \in \bm{\Omega} \setminus \bm{\Omega}_0,
\]
where $\bm{\theta}_0$ is the true parameter and $\bm{\Omega}_0$ is a subset of $\bm{\Omega}$. 
When $\bm{\Omega}_0$ is an $r$-dimensional subset of $\bm{\Omega}$, $\bm{\theta}_0$ is a boundary point of both $\bm{\Omega}_0$ and $\bm{\Omega}\setminus \bm{\Omega}_0$ but an interior point of $\bm{\Omega}$, under some regularity conditions, by the Wilk's theorem, the asymptotic distribution of the LRT  statistic, 
$\lambda_N := -2 \big(\underset{\bm{\theta}\in \bm{\Omega}_0}{\sup} l_N (\bm{\theta})  - \underset{\bm{\theta}\in \bm{\Omega}}{\sup} l_N (\bm{\theta}) \big)$, will be $\chi^2(p-r)$.
However, when $\bm{\theta}_0$ is a boundary point of $\bm{\Omega}$, the regularity condition is not satisfied and the conventional Chi-squared limiting distribution does not hold either.

In \cite{self1987asymptotic}, the authors studied the nonstandard tests where the parameter of the null model is on the boundary of the parameter space.
It is shown that when some of the true parameter values are on the boundary of the parameter space, under certain regularity conditions, the limiting distribution of the LRT statistic is the same as the distribution of the projection of the Gaussian random variable onto the region of admissible values for the mean.
Specifically, both the whole parameter space $\bm{\Omega}$ and the null parameter space $\bm{\Omega}_0$ are assumed to be regular enough to be approximated by cones with vertices at the true parameter $\bm{\theta}_0$, which is defined as below.
\begin{definition}
The set $\bm{\Omega} \subset \mathbb{R}^p$ is approximated at $\bm{\theta}_0$ by a cone with vertex at $\bm{\theta}_0$, $C_{\bm{\Omega}}$, if
\begin{align*}
    &(1)\inf_{\bm{x}\in C_{\bm{\Omega}}}||\bm{x}-\bm{y}|| = o(||\bm{y}-\bm{\theta}_0||), \quad \forall \bm{y}\in \bm{\Omega}, \\
    &(2)\ \inf_{\bm{y}\in \bm{\Omega}} ||\bm{x}-\bm{y}|| = o(||\bm{x}-\bm{\theta}_0||), \quad \forall \bm{x} \in C_{\bm{\Omega}}.
\end{align*}
\end{definition}

\noindent When the model is identifiable, with some further regularity conditions (see Section 1 in \cite{self1987asymptotic} for details), 
the following asymptotic distribution of the LRT statistic has been derived.

\begin{theorem} [Self and Liang, 1987]
Let $\bm{Z}$ be a random variable with a multivariate Gaussian distribution with mean $\bm{\theta}_0$ and covariance matrix $I^{-1}(\bm{\theta}_0)$, where $I(\bm{\theta}) = N^{-1}I_N(\bm{\theta})$ and $I_N(\bm{\theta})$ is the second derivative of the log-likelihood function $l_N(\bm{\theta})$.
Let $C_{\bm{\Omega}_0}$ and $C_{\bm{\Omega}}$ be non-empty cones approximating $\bm{\Omega}_0$ and $\bm{\Omega}$ at $\bm{\theta}_0$, respectively.
Then the asymptotic distribution of the likelihood ratio test statistic is the same as the distribution of the likelihood ratio test of $\bm{\theta}\in C_{\bm{\Omega}_0}$ versus the alternative $\bm{\theta} \in C_{\bm{\Omega}}$ based on a single realization $\bm{Z}$ when $\bm{\theta} = \bm{\theta}_0$.
\label{thm:lrt}
\end{theorem}

Following \cite{self1987asymptotic}, the asymptotic representation of the LRT statistic given by Theorem \ref{thm:lrt} can be written as 
\begin{equation}
	\sup_{\bm{\theta} \in C_{\bm{\Omega}}-\bm{\theta}_0}\{-(\bm{Z}-\bm{\theta})^\top I(\bm{\theta}_0)(\bm{Z}-\bm{\theta})\} \ - \sup_{\bm{\theta} \in C_{\bm{\Omega}_0} - \bm{\theta}_0}\{-(\bm{Z}-\bm{\theta})^\top I(\bm{\theta}_0)(\bm{Z}-\bm{\theta})\},
	\label{eq-self-1}
\end{equation}
where $\bm{Z}$ has a multivariate Gaussian distribution with mean $\bm{0}$ and covariance matrix $I^{-1}(\bm{\theta}_0)$. 
We can further rewrite it as
\begin{equation}
	\inf_{\bm{\theta} \in \Tilde{C}_0 }||\Tilde{\bm{Z}}-\bm{\theta}||^2 - \inf_{\bm{\theta} \in \Tilde{C}} ||\Tilde{\bm{Z}} - \bm{\theta}||^2,
	\label{eq-self-2}
\end{equation}
where $\Tilde{C} = \{\Tilde{\bm{\theta}}: \Tilde{\bm{\theta}} = \Lambda^{1/2}P^T\bm{\theta}, \ \forall\ \bm{\theta} \in C_{\bm{\Omega}}-\bm{\theta}_0 \}$,
$\Tilde{C_0} = \{\Tilde{\bm{\theta}}:\Tilde{\bm{\theta}} = \Lambda^{1/2}P^T\bm{\theta}, \ \forall\ \bm{\theta} \in C_{\bm{\Omega}_0}-\bm{\theta}_0\}$, 
$\Tilde{\bm{Z}}$ follows a multivariate Gaussian distribution with mean $\bm{0}$ and the identity covariance matrix,
and $P\Lambda P^T$ represents the spectral decomposition of $I(\bm{\theta}_0)$. 
Therefore, after the orthogonal transformation, the distribution in equation \eqref{eq-self-1} can be computed using the standard Gaussian distribution.

This result provides a promising direction for the hypothesis testing in the hierarchical CDM setting, and we consider a simple example in Example \ref{ex:test}.

\begin{example}
Consider a DINA model with two latent attributes.
Suppose that we want to test whether the first attribute is a prerequisite for the second attribute, that is, the hierarchical structure $\mathcal{E}_0 = \{1\rightarrow 2\}$.
Assume that the identifiability conditions in Proposition \ref{prop-dina} are satisfied.
The model parameters include the proportion parameters and item parameters $\{p_{(0,0)},\ p_{(0,1)},\ p_{(1,0)},\ p_{(1,1)},\\ \ s_j,\ g_j, \ j = 1,\dots,J\}$,
so the total number of parameters is $3+2\times J$, noting that the proportion parameter vector $\bm{p} = (p_{(0,0)},\ p_{(0,1)},\ p_{(1,0)},\ p_{(1,1)})$ lies in the 3-simplex.
To test the hierarchy $\mathcal{E}_0$, it is equivalent to test
\[
\text{H}_0: p_{(0,1)} = 0 \quad \text{vs} \quad  \text{H}_1: p_{(0,1)}\neq 0. 
\]
Therefore we have one parameter of interest that has true value on the boundary and $2+2\times J$ nuisance parameters with true values not on the boundary.
After an orthogonal transformation, we have $\Tilde{C} = [0, \infty)\times \mathbb{R}^{2+2\times J}$ and $\Tilde{C_0} = \{0\}\times \mathbb{R}^{2+2\times J}$ and thus the asymptotic distribution of the LRT statistic is reduced to
\[
\Tilde{Z}_1^2\cdot I(\Tilde{Z}_1>0),
\]
where $\Tilde{Z}_1$ follows a standard univariate gaussian distribution.
Therefore the limiting distribution of the LRT statistic is a mixture of Chi-squared distribution $\frac{1}{2}\chi_0^2 + \frac{1}{2}\chi_1^2$.
\label{ex:test}
\end{example}

\begin{figure}[ht]
    \centering
    \subfigure{
        \includegraphics[width=2in]{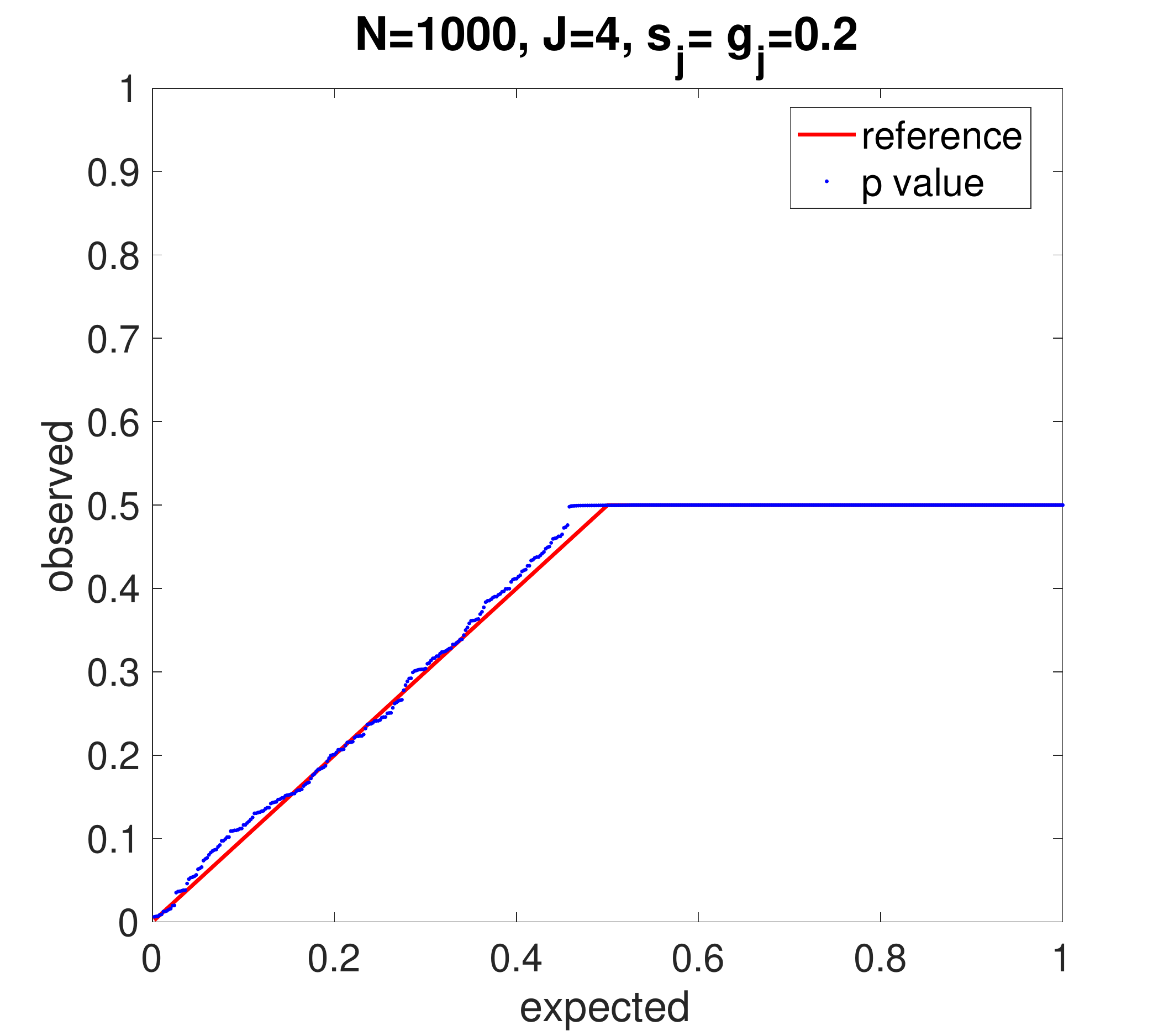}}
    \hspace{0in}
    \subfigure{
        \includegraphics[width=2in]{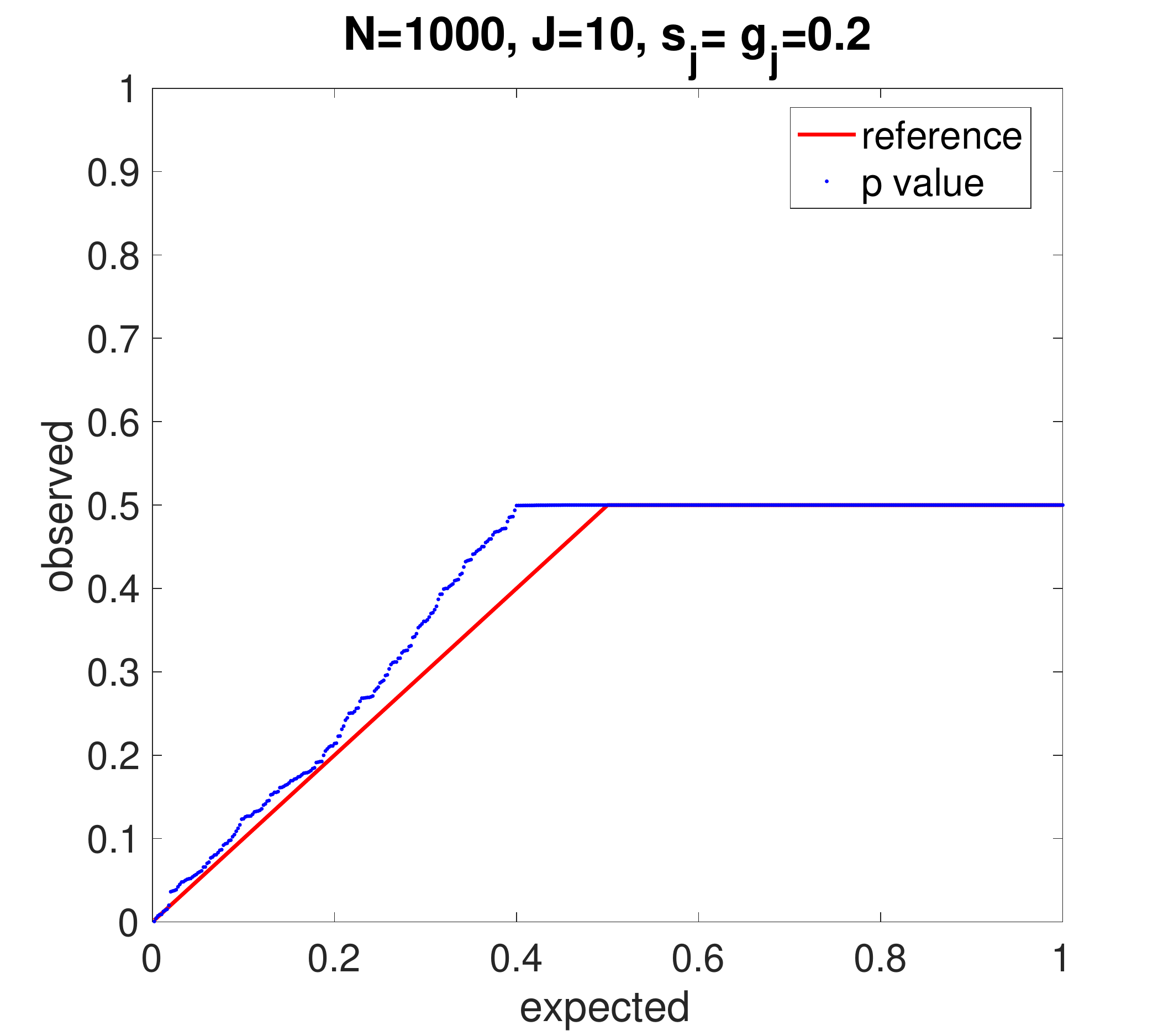}}
    \hspace{0in}
    \subfigure{
        \includegraphics[width=2in]{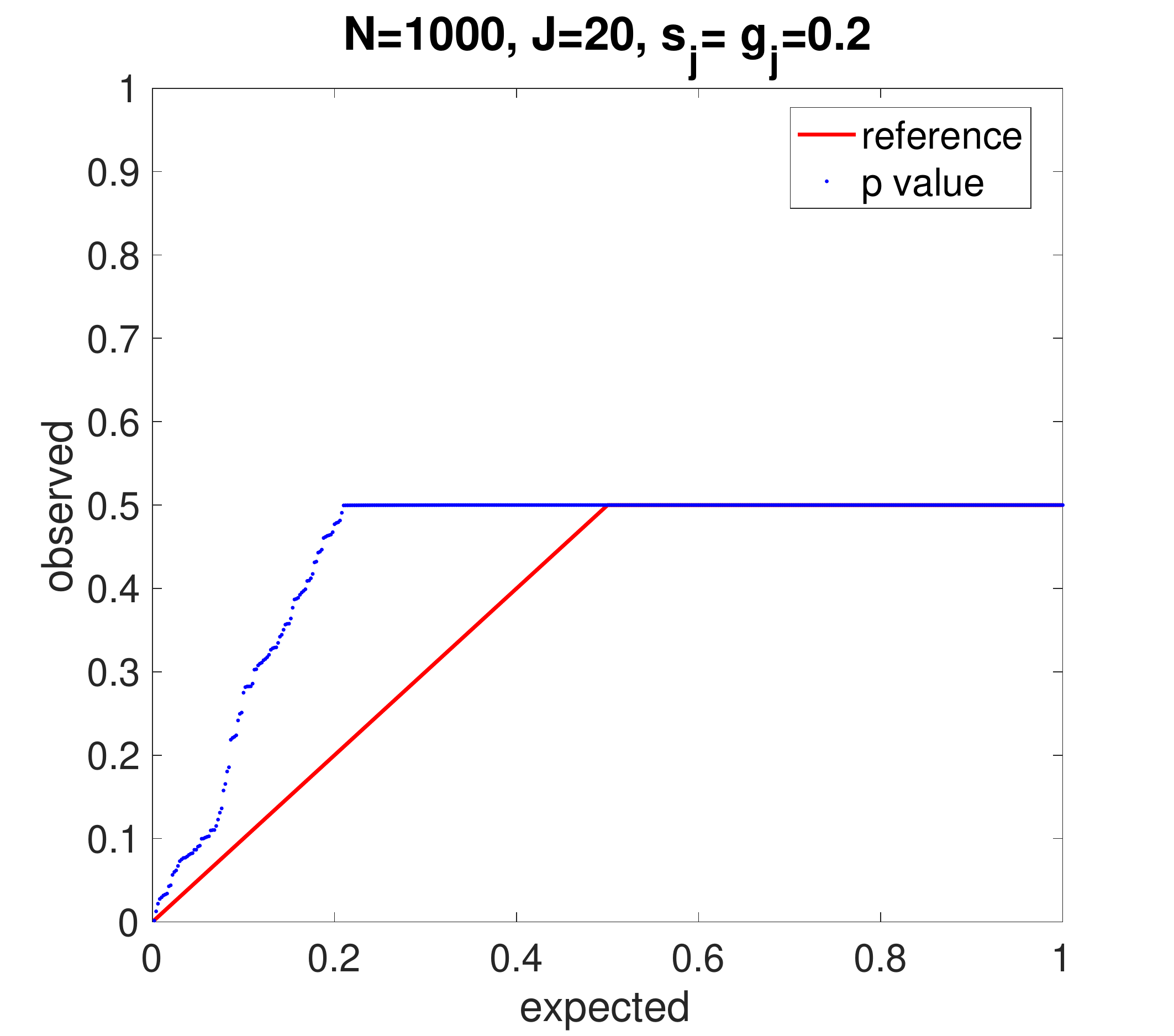}}
    \\
    \subfigure{
        \includegraphics[width=2in]{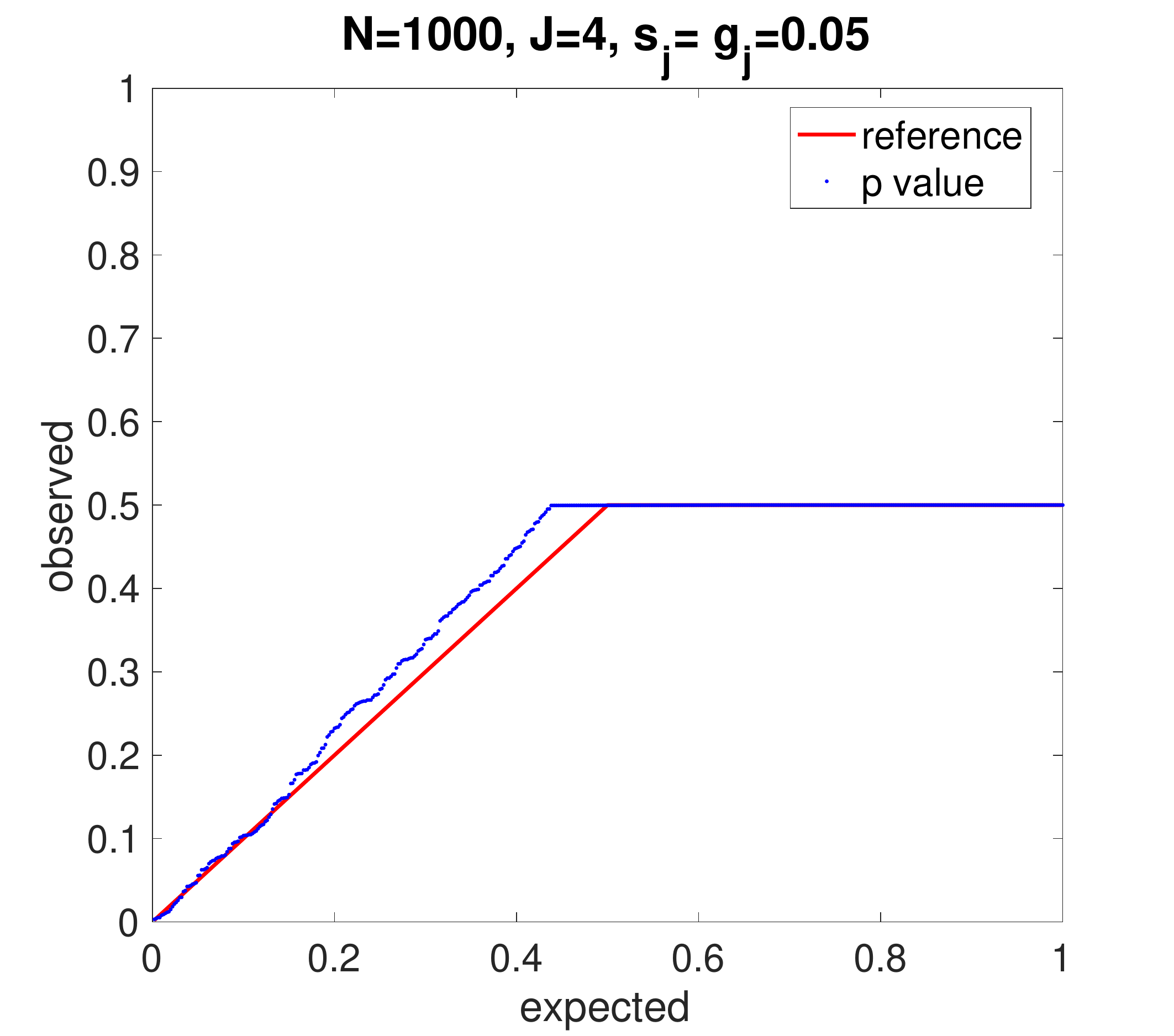}}
    \hspace{0in}
    \subfigure{
        \includegraphics[width=2in]{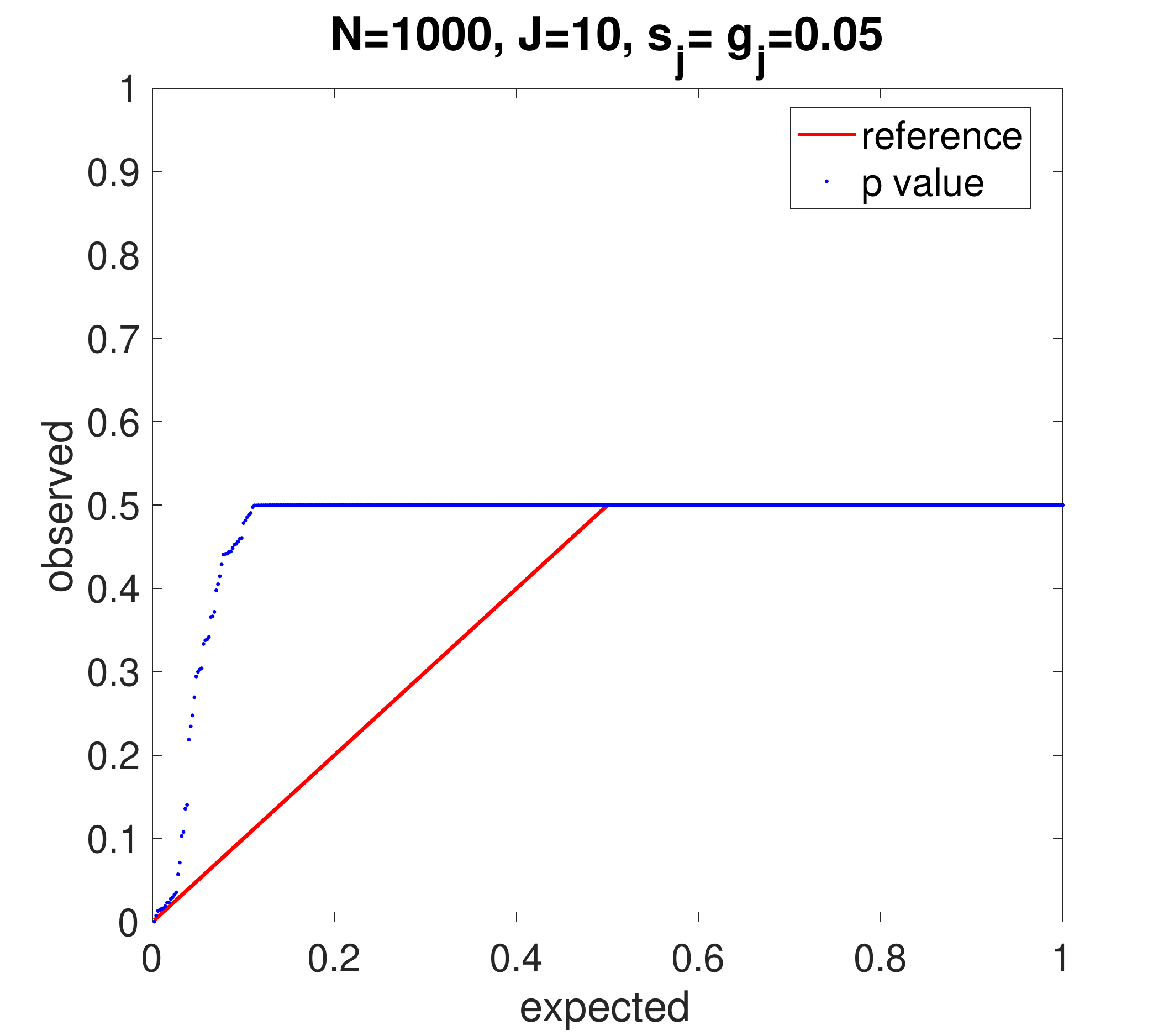}}
    \hspace{0in}
    \subfigure{
        \includegraphics[width=2in]{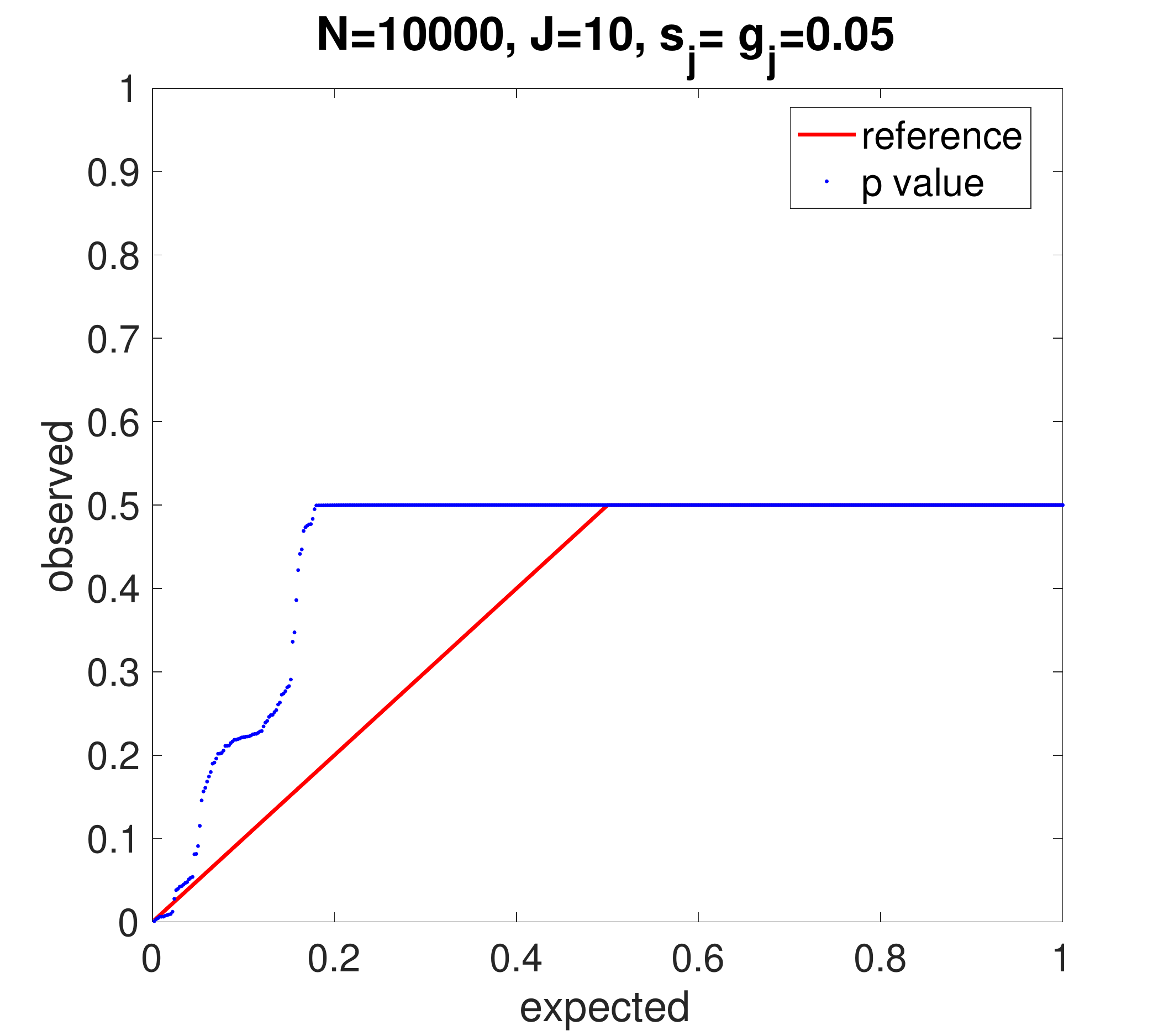}}
    
    \caption{QQ-plots for Example \ref{ex:test} under various settings. The x-axis is the expected percentile of the p-values under the null hypothesis, and the y-axis is the percentile of the observed p-values. The observed p-values are plotted by blue points, and the p-values for the reference limiting distribution $\frac{1}{2}\chi_0^2 + \frac{1}{2}\chi_1^2$ are plotted as red lines. If the blue points are close to the red lines, it indicates that the empirical distribution of the observed p-values approximates the asymptotic distribution well.
    }
    \label{fig:test}  
\end{figure}

In Example \ref{ex:test}, we derive the closed form of the limiting distribution of the LRT statistic in the DINA model with two latent attributes and a linear hierarchy.
In this example, we take the advantage of the fact that there is only one boundary parameter and it occurs as the parameter of interest.
However, the asymptotic distribution of the LRT statistic in fact becomes considerably more complicated if there are more latent attributes and more complex hierarchical structures.
Moreover, even in the simple setting as in Example \ref{ex:test}, the convergence may be very slow if the number of items $J$ is large or the guessing parameter $g_j$ and slipping parameter $s_j$ are close to the boundary, as  illustrated in Figure \ref{fig:test}.
Specifically, in Figure \ref{fig:test}, we present the p-values under various settings for Example \ref{ex:test}.
The observed p-values are plotted by blue points, and the p-values for the reference distribution $\frac{1}{2}\chi_0^2 + \frac{1}{2}\chi_1^2$ are plotted as the red lines.
The first row in Figure \ref{fig:test} contains three plots of p-values with the same sample size and item parameters but different numbers of items.
It is noted that when the number of items was small, the observed p-values were very close to those of the mixture Chi-squared limiting distribution.
However, as the number of items increased, the gap between the observed p-values and the reference limiting distribution became larger.
The second row in Figure \ref{fig:test} contains three plots of p-values with more extreme item parameters.
Compared with the plots in the first row, it is shown that when the item parameters were close to the boundary, the convergence of the LRT statistic became much slower, and such testing tended to fail even with a large sample size $N=10,000$.

As pointed out in \cite{self1987asymptotic}, even though based on Theorem \ref{thm:lrt} we can derive the asymptotic distribution of the LRT statistic for any fixed $\bm{\theta}_0\in\bm{\Omega}_0$, this distribution is generally different for different $\bm{\theta}_0$.
Moreover, these distributions typically vary over $\bm{\Omega}_0$ in a discontinuous way when some of the nuisance parameters may also be on the boundary.
This discontinuity can affect the quality of the asymptotic approximation much.
As in our example, when the slipping parameter $s_j$ and the guessing parameter $g_j$ in the DINA model are close to the boundary, the distribution with a given finite sample may be far away from the weighted Chi-squared mixture as described in Example \ref{ex:test}.

We provide further insights on why the convergence was slow when the number of items was large or the item parameters were close to the boundaries as shown in Figure \ref{fig:test}.
The average log-likelihood of CDMs is given by 
\[
l_N / N := \frac{1}{N}\sum_{i=1}^N \log \Big(\sum_{\bm{\alpha}}p_{\bm{\alpha}} P(\bm{R}_i\mid \bm{\alpha})\Big).
\]
Consider Example \ref{ex:test} where we are interested in the hierarchy $\{ 1\xrightarrow{}2 \}$ and want to test whether $p_{(0,1)}=0$.
If we write $p_{(0,0)}=1-\sum_{\bm{\alpha}\neq (0,0)}p_{\bm{\alpha}}$, then 
\[
l_N / N = \frac{1}{N}\sum_{i=1}^N \log \Big((1-\sum_{\bm{\alpha}\neq (0,0)}p_{\bm{\alpha}})P\big(\bm{R}_i \mid \bm{\alpha}=(0,0)\big)+\sum_{\bm{\alpha}\neq (0,0)}p_{\bm{\alpha}}P(\bm{R}_i \mid \bm{\alpha})\Big).
\]
The derivative of the log-likelihood w.r.t. $p_{(0,1)}$ becomes
\begin{align}
	\nonumber
    \frac{\partial l_N / N }{\partial p_{(0,1)}}&=\frac{1}{N}\sum_{i=1}^N \frac{P\big(\bm{R}_i\mid\bm{\alpha}=(0,1)\big)-P\big(\bm{R}_i\mid\bm{\alpha}=(0,0)\big)}{\sum_{\bm{\alpha}} p_{\bm{\alpha}} P\big(\bm{R}_i\mid\bm{\alpha}\big)}\\
    \nonumber
    &=\frac{1}{N}\sum_{i=1}^N \frac{P\big(\bm{R}_i\mid\bm{\alpha}=(0,1)\big)-P\big(\bm{R}_i\mid\bm{\alpha}=(0,0)\big)}{P(\bm{R}_i)}\\
    \label{eq-derivative}
&=\frac{1}{N}\sum_{i=1}^N
\sum_{\bm{r} \in\{0,1\}^J}I(\bm{R}_i = \bm{r})\frac{P\big(\bm{r}\mid\bm{\alpha}=(0,1)\big)-P\big(\bm{r} \mid\bm{\alpha}=(0,0)\big)}{P(\bm{r})}.
\end{align}
When the null hypothesis that $p_{(0,1)}=0$ is true, by the strong law of large number, we have 
\begin{align*}
    \frac{\partial l_N / N }{\partial p_{(0,1)}}\Big|_{p_{(0,1)}=0}  \overset{a.s.}{\longrightarrow} \quad &\mathbb{E}_0\Big[
\sum_{\bm{r} \in\{0,1\}^J}I(\bm{R} = \bm{r})\frac{P\big(\bm{r} \mid\bm{\alpha}=(0,1)\big)-P\big(\bm{r} \mid\bm{\alpha}=(0,0)\big)}{P(\bm{r} )}\Big]\\
    &= \sum_{\bm{r} \in \{0,1\}^J} \Big(P\big(\bm{r}\mid\bm{\alpha}=(0,1)\big)- P\big(\bm{r} \mid\bm{\alpha}=(0,0)\big)\Big)\\
    &= \sum_{\bm{r} \in \{0,1\}^J} P\big(\bm{r} \mid\bm{\alpha}=(0,1)\big) - \sum_{\bm{r} \in \{0,1\}^J} P\big(\bm{r} \mid\bm{\alpha}=(0,0)\big) \\
    &= 0.
\end{align*}
However, since the number of possible response patterns is $|\{0,1\}^J| = 2^J$ which grows exponentially with the number of items $J$, it requires an exponentially growing sample size to cover all the possible response patterns, and therefore the convergence can be slow when $J$ is large.

Next, consider the case when the item parameters are close to the boundary. 
Note that
\begin{align*}
    P\big(\bm{R}\mid\bm{\alpha}\big) = \prod_{j=1}^J P\big(R_j \mid \bm{\alpha}\big)
	= \prod_{j=1}^J \Big(g_j^{1-\Gamma_{j, \alpha}}(1 - s_j)^{\Gamma_{j, \alpha}}\Big)^{R_{j}}\Big((1 - g_j)^{1-\Gamma_{j, \alpha}}s_j^{\Gamma_{j, \alpha}}\Big)^{1-R_{j}}.
\end{align*}
When the item parameters are very close to the boundaries, that is, $s_j$ and $g_j$ are very close to 0, the model becomes near deterministic.
For simplicity, let $s_j = g_j = \delta$ which is very close to 0 for all $j\in [J]$.
Then 
\begin{align*}
	P(\bm{R}  = \bm{r} \mid \bm{\alpha}) & = \prod_{j=1}^J \Big(\delta^{(1-\Gamma_{j, \bm{\alpha}})}(1 - \delta)^{\Gamma_{j, \bm{\alpha}}}\Big)^{r_{j}}\Big((1 - \delta)^{(1-\Gamma_{j, \bm{\alpha}})}\delta^{\Gamma_{j, \bm{\alpha}}}\Big)^{1-r_{j}}\\
	& = \prod_{r_j = 1} \delta^{(1 - \Gamma_{j, \bm{\alpha}})}(1-\delta)^{\Gamma_{j, \bm{\alpha}}} \prod_{r_j = 0} (1-\delta)^{(1-\Gamma_{j, \bm{\alpha}})} \delta^{\Gamma_{j, \bm{\alpha}}}\\
	& = \delta^{\sum_{r_j = 1}(1 - \Gamma_{j, \bm{\alpha}} ) + \sum_{r_j = 0} \Gamma_{j, \bm{\alpha}} } \cdot ( 1-\delta)^{\sum_{r_j = 1}\Gamma_{j, \bm{\alpha}} + \sum_{r_j = 0} (1 - \Gamma_{j, \bm{\alpha}})}.
\end{align*}
For $\bm{r} = \bm{\Gamma}_{\cdot, \bm{\alpha}}$, we have $P(\bm{R}  = \bm{\Gamma}_{\cdot, \bm{\alpha}} \mid \bm{\alpha}) = (1-\delta)^J$.
And for any $\bm{r} \neq \bm{\Gamma}_{\cdot, \bm{\alpha}}$, $ \delta^J \leq P(\bm{R}  = \bm{r} \mid \bm{\alpha}) \leq \delta$.
Moreover, when $p_{(0,1)} = 0$, since $P(\bm{R}  = \bm{r}) = \sum_{\bm{\alpha} \neq (0, 1)} p_{\bm{\alpha}} P(\bm{R} = \bm{r} \mid \bm{\alpha})$, we have 
\[
P(\bm{R}  = \bm{\Gamma}_{\cdot, \bm{\alpha}}) \geq p_{\bm{\alpha}}(1-\delta)^J, \text{ for } \bm{\alpha} \in \mathcal{A} = \{(0,0), (1,0), (1,1)\},
\]
and
\[
P\big(\bm{R} \neq \bm{\Gamma}_{\cdot, \bm{\alpha}}, \  \bm{\alpha} \in \mathcal{A}\big) \leq 1 - (1 - \delta)^J \rightarrow 0 \text{ as } \delta \rightarrow 0.
\]
Therefore from the above discussions, when $p_{(0,1)} = 0$, the probability mass is concentrated around three response patterns $\bm{\Gamma}_{\cdot, \bm{\alpha}}$ for $\bm{\alpha} \in \mathcal{A} = \{(0, 0), (1, 0), (1, 1)\}$.
For the terms in the RHS of \eqref{eq-derivative}, when $\bm{r} = \Gamma_{\cdot, (0,0)}$, we have 
\begin{align*}
& \quad \big[P\big(\bm{R} = \bm{r} \mid\bm{\alpha}=(0,1)\big)-P\big(\bm{R} = \bm{r}\mid\bm{\alpha}=(0,0)\big)\big]\big(P(\bm{R} = \bm{r})\big)^{-1} 
\\	& \in \Big[\frac{\delta^J - (1-\delta)^J}{(1-\delta)^J p_{(0,0)}},  \frac{\delta - (1-\delta)^J}{(1-\delta)^J p_{(0,0)} + \delta(1-p_{(0,0)})}\Big] \longrightarrow - 1/p_{(0,0)} \text{ as } \delta \rightarrow 0.	
\end{align*}

When $\bm{r} = \Gamma_{\cdot, (1,0)}$ (or $\Gamma_{\cdot, (1,1)}$), we have 
\begin{align*}
	& \quad\big[P\big(\bm{R} = \bm{r} \mid\bm{\alpha}=(0,1)\big)-P\big(\bm{R} = \bm{r}\mid\bm{\alpha}=(0,0)\big)\big]\big(P(\bm{R} = \bm{r})\big)^{-1} \\
	& \in \Big[\frac{\delta^J - \delta}{(1-\delta)^J p_{(1,0)}},  \frac{\delta - \delta^J}{(1-\delta)^J p_{(1,0)} + \delta(1-p_{(1,0)})}\Big] \longrightarrow 0 \text{ as } \delta \rightarrow 0.
\end{align*}
Therefore the terms in the RHS of \eqref{eq-derivative} also concentrate around two points, $-1/ p_{(0,0)}$ and 0, making the convergence slow since more data points are needed to have it converge to 0.

Based on the above discussions about the asymptotic behaviors of the LRT statistic under the nonstandard conditions and in the hierarchical CDM setting, it has been shown that even in the simple setting where we could derive a closed form of the limiting distribution, the convergence can be very slow.
Moreover, the asymptotic distribution of the LRT will be much more complicated if we have more latent attributes and more complex hierarchical structures.
Therefore, it is not practical to use the theoretical limiting distribution of the LRT statistic to test latent hierarchical structures in CDMs, especially considering that the number of test items is usually relatively large (e.g. more than 20).

\subsection{Bootstrap and Numerical Studies}
\label{sec-boot}

From the discussions about the LRT in Section \ref{sec-limiting}, we learn that the limiting distribution of the LRT statistic under latent hierarchical structures can be very complicated and the convergence can be slow when the number of items is large or the item parameters are close to the boundary even in simple settings.  
To overcome these difficulties, we propose to use the bootstrap method as an alternative to the asymptotic limiting distribution method.
The bootstrap method  \citep{efron1979bootstrap} has been shown to be successful in many nonstandard situations. 
The basic idea of bootstrap is treating inference of the true probability distribution, given the original data, as being analogous to the inference of the empirical distribution, given the resampled data. 
If the empirical distribution is a reasonable approximation to the true distribution, then the bootstrap method will provide good inferences.

In this section, we consider two different bootstrap procedures: nonparametric bootstrap and parametric bootstrap.
The idea of nonparametric bootstrap is to simulate data from the empirical distribution by directly resampling from the original data.
To be specific, in nonparametric bootstrap, we draw samples of the same size from the original data with replacement.
Then the statistic of interest is computed based on the resampled data set and we repeat this routine many times. 
The steps for nonparametric bootstrap are summarized as below:
\begin{enumerate}
\item[Step 1.] Initially estimate the model with the specified hierarchy under the null hypothesis, and the model  under the alternative hypothesis (without the null hypothesis hierarchy constraints), and calculate the LRT statistic.
\item[Step 2.] Draw a sample of the same size with replacement from the original data and calculate the LRT statistic.
\item[Step 3.] Repeat Step 2 independently many times and estimate the distribution of the LRT statistic.
\item[Step 4.] Estimate the p-value by comparing the distribution obtained in Step 3 with the LRT statistic obtained in Step 1. 
	Then this p-value is used to determine whether the null model with the specified null hierarchy should be rejected in favor of the model without the hierarchical constraints.
\end{enumerate}

The idea of parametric bootstrap is to simulate data based on good estimates of distribution parameters, often by maximum likelihood.
In parametric bootstrap, a parametric model is fitted to the original data, and samples are drawn from this fitted model. 
The steps for parametric bootstrap  are similar to those of nonparametric bootstrap except for Step 2:

\begin{enumerate}
\item [Step 2*.] Based on the estimates of the model with specified hierarchy from step 1, generate a bootstrap sample from the fitted model and calculate the LRT statistic.
\end{enumerate}

Next we conduct comprehensive simulation studies to compare parametric bootstrap and nonparametric bootstrap for testing latent hierarchical structures under various settings.
We considered four different hierarchical structures shown in Figure \ref{fig:hier}.
For the data generating process, we considered the DINA model and the GDINA model respectively.
For both models we included three different sample sizes ($N$ = 200, 500, or 1000) and the number of items was set to 30 ($J = 30$).
In terms of uncertainty, two levels of guessing and slipping parameters in the DINA model were included ($s_j = g_j = 0.1$ or 0.2 for $j \in [J]$).
For the GDINA model, we also considered two different uncertainty levels, where the highest item parameter was $0.9$ or $0.8$, and the lowest item parameter was $0.1$ or $0.2$.
The other item parameters in between were equally spaced.
To satisfy testability conditions, the $Q$-matrix contained two identity sub-matrices and the remaining items were randomly generated.
For each scenario, we performed 500 independent repetitions and in each repetition we generated bootstrap samples for 500 times.
To fit the models under the null and alternative hypotheses, we used R package ``CDM".

\begin{figure}[H]
    \centering
    \subfigure[DINA, $\theta_j^+ = 0.9, \theta_j^- = 0.1$]{
    \includegraphics[width=3.1in]{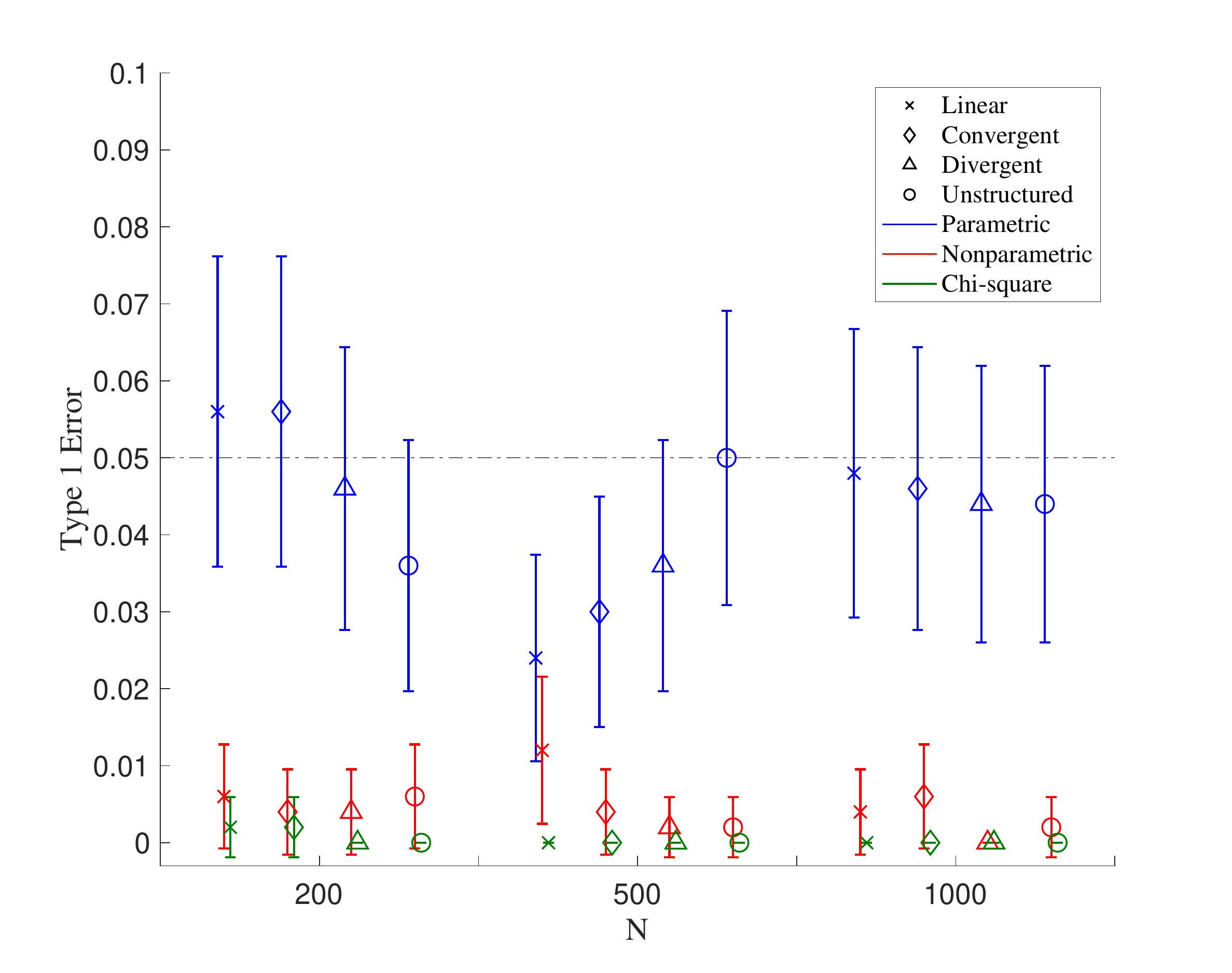}
    }
    \subfigure[DINA, $\theta_j^+ = 0.8, \theta_j^- = 0.2$]{
    \includegraphics[width=3.1in]{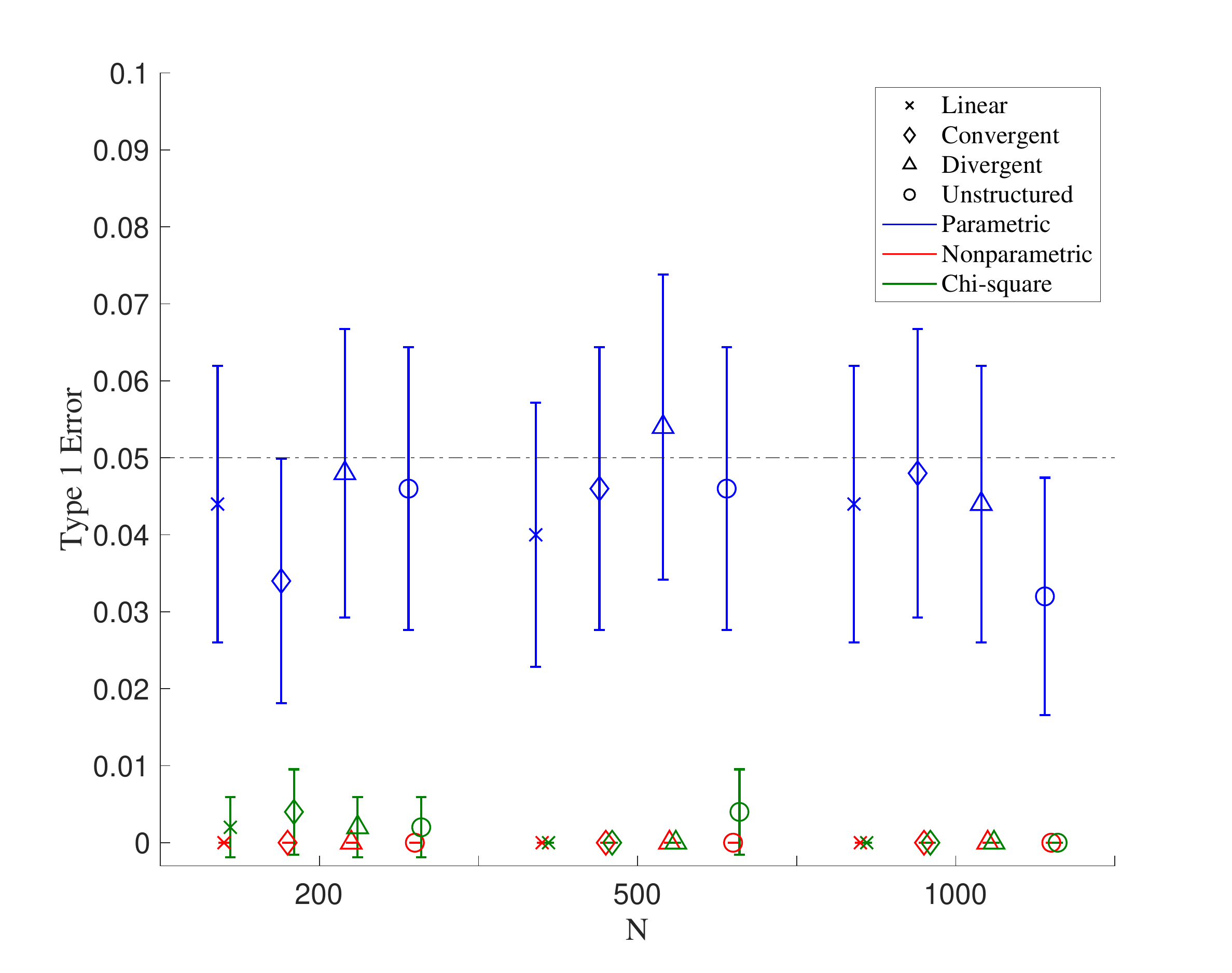}
    }
	\\
	\subfigure[GDINA, $\theta_j^+ = 0.9, \theta_j^- = 0.1$]{
    \includegraphics[width=3.1in]{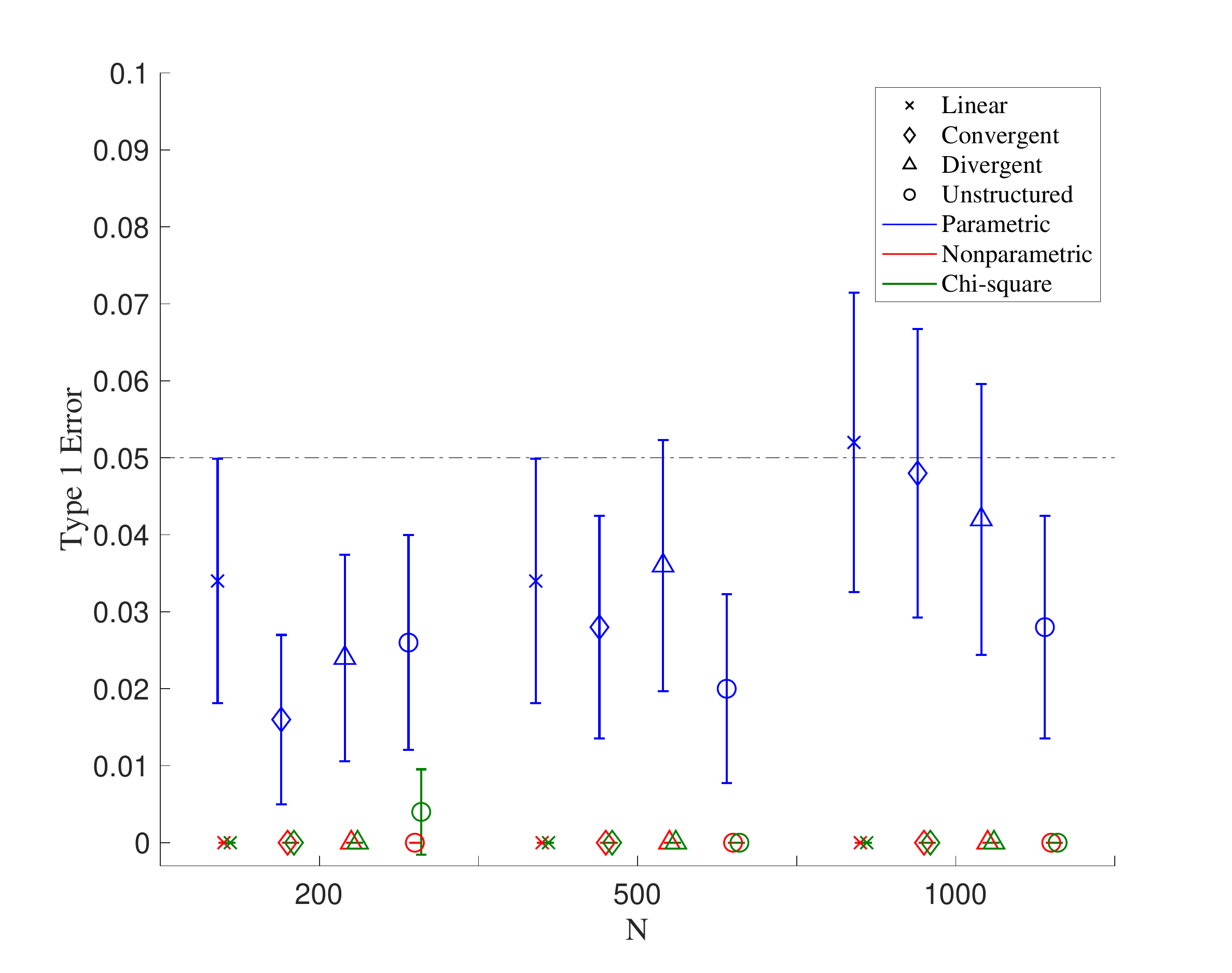}
    }
    \subfigure[GDINA, $\theta_j^+ = 0.8, \theta_j^- = 0.2$]{
    \includegraphics[width=3.1in]{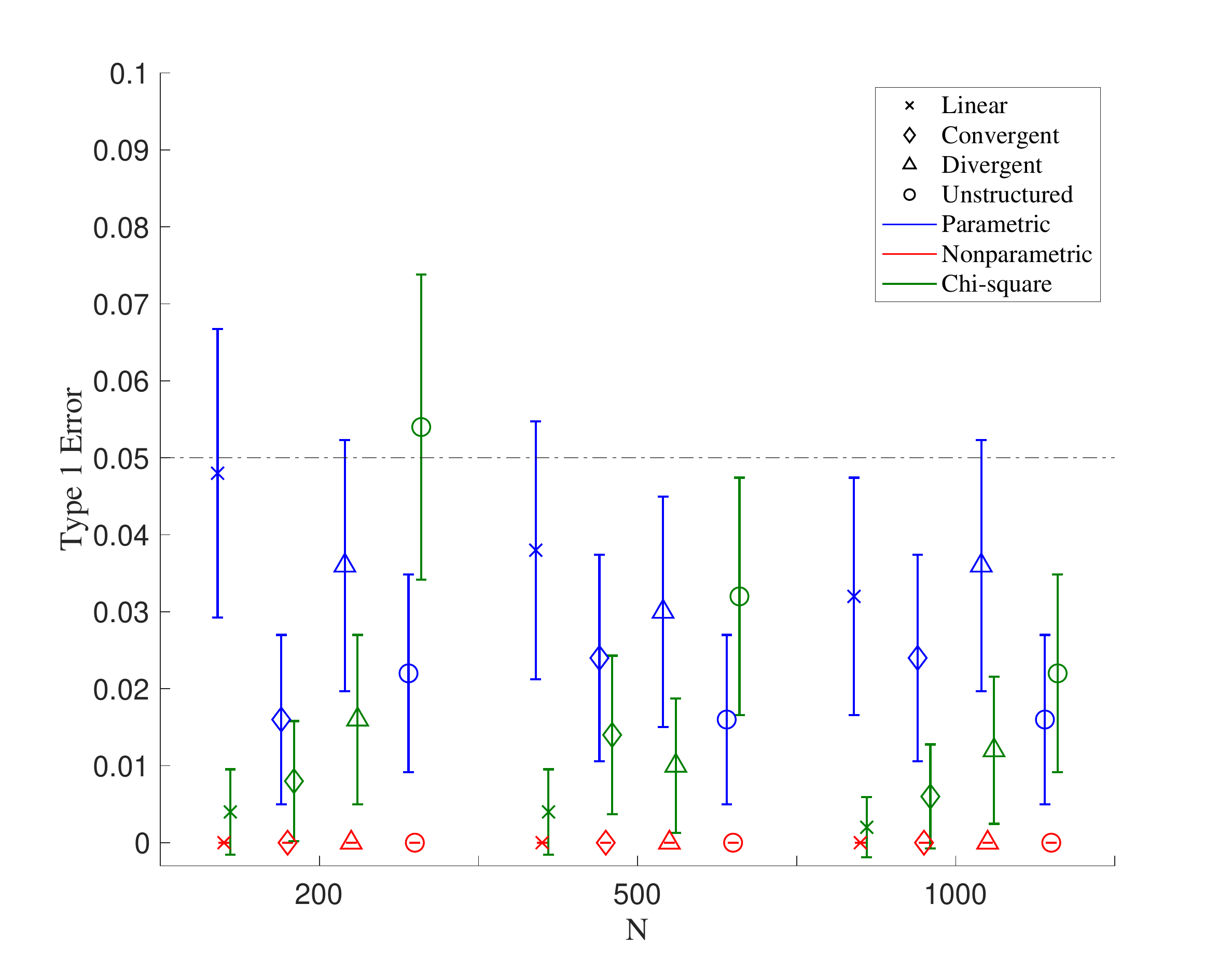}
    }	
    
    \caption{Type 1 Errors. Different colors indicate different testing procedures. Different marker shapes stand for different hierarchical structures. The middle points are the means of the type I errors and the vertical errors bars  with $\pm$2 s.e. are constructed based on  500 replications.
    $\theta_j^+ = 0.9,\ \theta_j^- = 0.1$ corresponds to the case with low noises, and $\theta_j^+ = 0.8, \ \theta_j^- = 0.2$ corresponds to the case with high noises.}
    \label{fig:type1}
\end{figure}

The type I errors with significance level $\alpha = 0.05$ under different settings are plotted in Figure \ref{fig:type1}.
The corresponding error bars are plotted for uncertainty quantification of the Monte Carlo errors.
We also include the na\"ive Chi-squared test for comprehensive comparison.
From the plots, we can see that the type I errors for parametric bootstrap were around 0.05 in most cases and therefore parametric bootstrap controlled the type I errors generally well.
By contrast, nonparametric bootstrap was too conservative and the type I errors for nonparametric bootstrap were very close to 0.
In terms of the na\"ive Chi-squared test, it was also very conservative in most cases even though the type I errors for the GDINA model with larger noises under the ``unstructured" hierarchy were closer to the significance level 0.05.

\begin{figure}[H]
    \centering
    \subfigure{
    \includegraphics[width=6.5in]{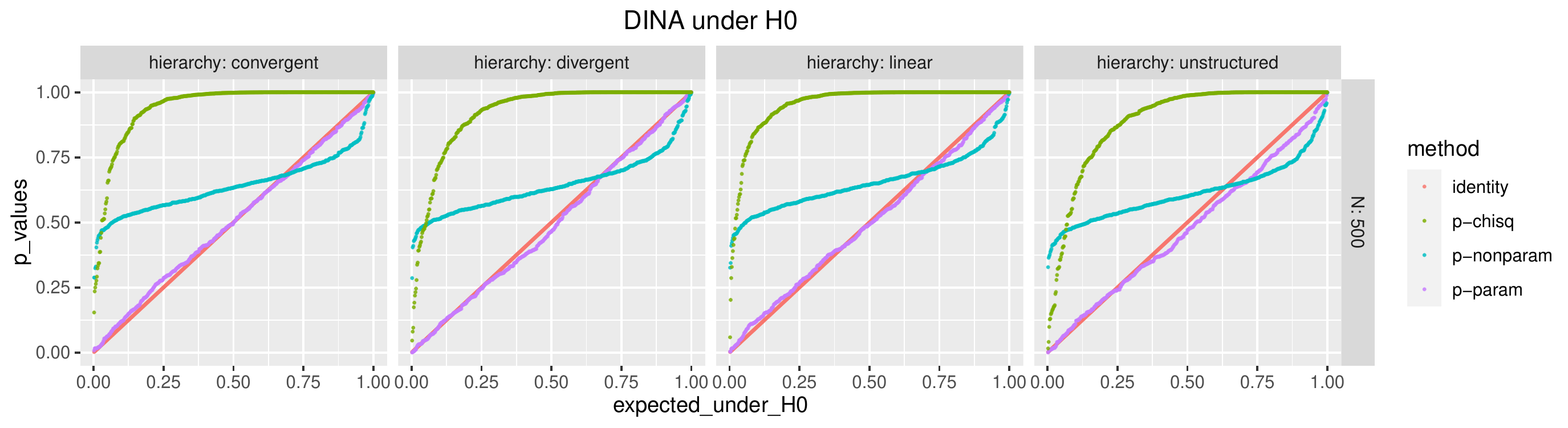}
    }
    \caption{QQ-plots for the p-values of the DINA model under the null hypothesis where $\theta_j^+ = 0.8,\ \theta_j^- = 0.2$ that corresponds to the case with high noises.. 
    }
    \label{fig:dina-qq}
\end{figure}

\begin{figure}[H]
    \centering
    \subfigure{
    \includegraphics[width=6.5in]{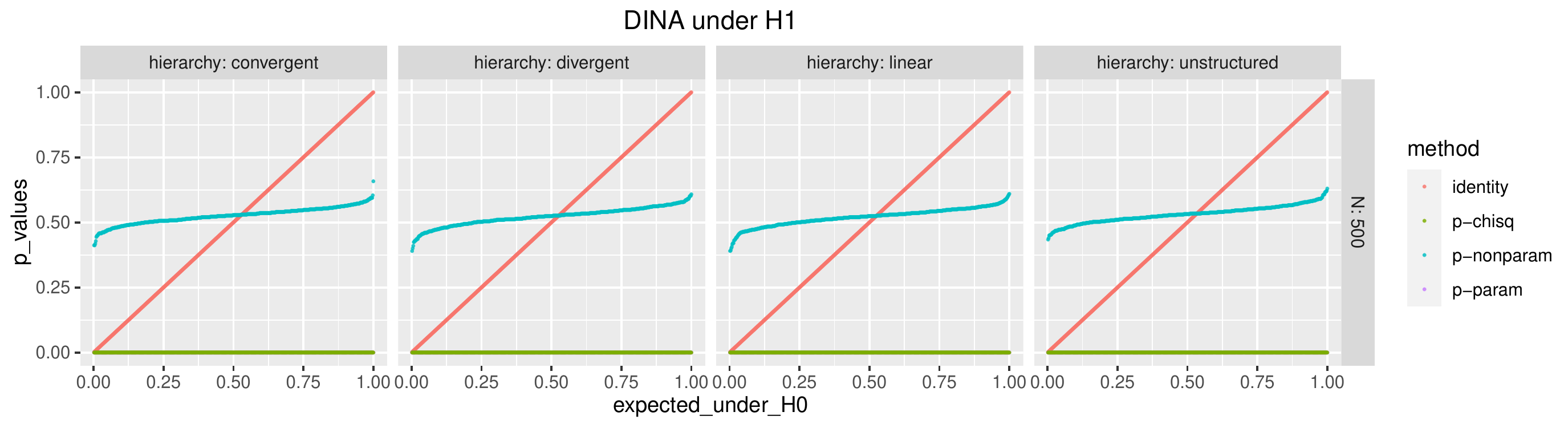}
    }
    \caption{QQ-plots for the p-values of the DINA model under the alternative hypothesis where $\theta_j^+ = 0.8,\ \theta_j^- = 0.2$ corresponding to high noises.    The expected quantiles are the expected quantiles of the p-values under the null hypothesis, that is, the uniform distribution on [0,1].
    }
    \label{fig:dina-qq-power}
\end{figure}

\begin{figure}[H]
    \subfigure{
    \includegraphics[width=6.5in]{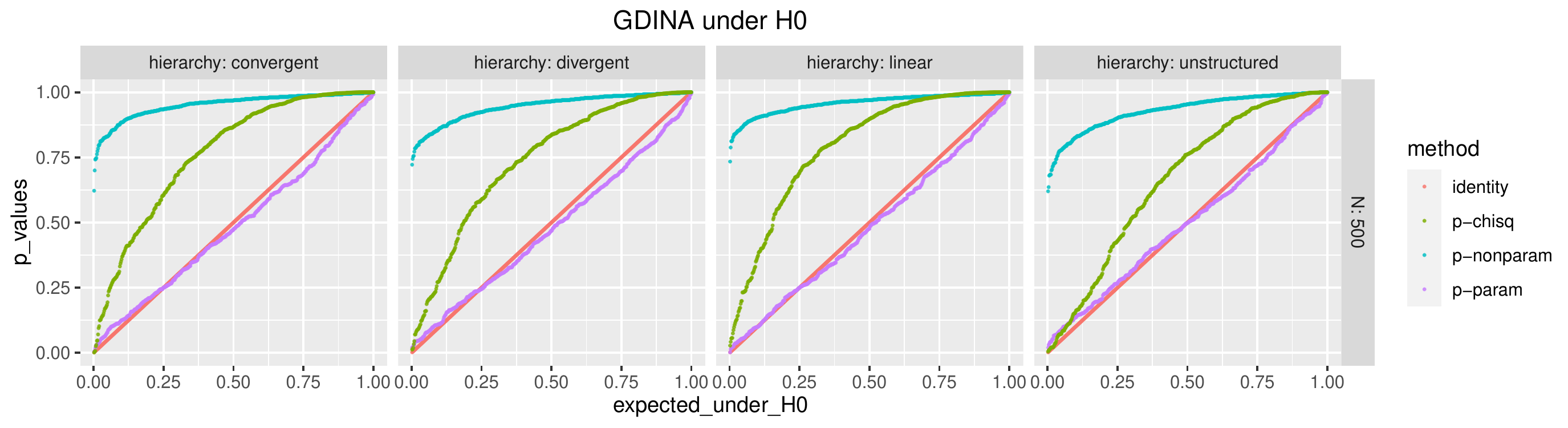}
    }
    \caption{QQ-plots for the p-values of the GDINA model under the null hypothesis where $\theta_j^+ = 0.8,\ \theta_j^- = 0.2$ corresponding to the case with high noises.
    }
    \label{fig:gdina-qq}
\end{figure}

\begin{figure}[H]
    \centering
    \subfigure{
    \includegraphics[width=6.5in]{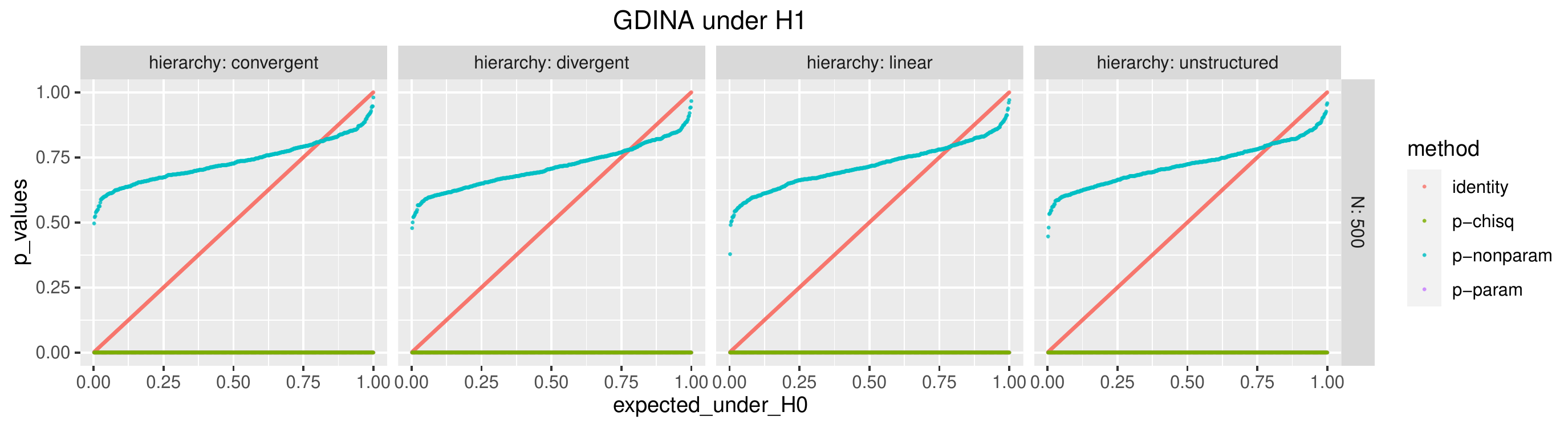}
    }
    \caption{QQ-plots for the p-values of the GDINA model under the alternative hypothesis where $\theta_j^+ = 0.8,\ \theta_j^- = 0.2$ corresponding to high noises.    The expected quantiles are the expected quantiles of the p-values under the null hypothesis, that is, the uniform distribution on [0,1].
    }
    \label{fig:gdina-qq-power}
\end{figure}

To further examine the behaviors of the testing procedures, the QQ plots for p-values under the null hypothesis are provided in Figure \ref{fig:dina-qq} and Figure \ref{fig:gdina-qq}.
For presentation brevity, we only present results of four hierarchies with the same noise level and sample size.
More comprehensive simulation results are presented in the supplementary material.
It is known that under the null hypothesis, the p-values should follow a uniform distribution on $[0,1]$.
In the QQ-plots, if the points are lying closer to the identity line, it indicates that it approximates the uniform distribution better. 
From Figure \ref{fig:dina-qq} and \ref{fig:gdina-qq}, one can see that under the null hypothesis, the p-values of parametric bootstrap approximated the uniform distribution on $[0,1]$ very well in almost all the settings.
By contrast, the p-values of nonparametric bootstrap and the na\"ive Chi-squared test were far away from the uniform distribution, indicating that these testing procedures are not reliable.
We also conducted power analysis where all the latent attribute profiles existed in the data generation process.
The true proportion parameters were equally assigned.
The QQ plots for p-values under the alternative hypothesis are shown in Figure \ref{fig:dina-qq-power} and Figure \ref{fig:gdina-qq-power} respectively.
To have more power, we expect the p-values to be small so that we would reject the null hypothesis.
Therefore in the QQ-plots, the closer to 0 the points are, the more powerful the test is.
From Figure \ref{fig:dina-qq-power} and \ref{fig:gdina-qq-power}, one can see that the p-values of parametric bootstrap and the na\"ive Chi-squared test were almost 0 and therefore the power was close to 1.
However, the p-values of nonparametric bootstrap were close or above 0.5, which means we would not reject the null hypothesis, making the power almost 0.
Taking both the type I error and power into consideration, the parametric bootstrap outperformed the other two testing procedures.

In order for nonparametric bootstrap to work, the empirical distribution of the sample data should be close to the true distribution, which may not hold for the cases here especially when the number of items is relatively large.
As we discussed in Section \ref{sec-limiting}, the total number of possible response patterns $2^J$ grows exponentially with the number of items $J$.
Therefore, in nonparametric bootstrap, we need a very large sample size to cover all the possible response patterns, which may explain the failures of nonparametric bootstrap.
On the contrary, in parametric bootstrap, we first fit a model from the original data and resample data from the fitted model, which incorporates the variance in data generation better and thus makes parametric bootstrap perform better.
Note that in \cite{templin2014hierarchical}, using the na\"ive Chi-squared test, the authors concluded that ``\textit{the DINA model cannot detect attribute hierarchies}".
However, through our comprehensive simulation, by using parametric bootstrap, the attribute hierarchies can also be well detected in the DINA model.

\medskip
In summary, based on our discussions about the failure of the limiting distribution of LRT in Section \ref{sec-limiting} and the simulation results in Section \ref{sec-boot}, we recommend using parametric bootstrap to perform hypothesis testing for latent hierarchical structures in CDMs.

{}

{}

{}

{}

\section{Real Data Analysis}
\label{sec-real}

In this section, we perform hypothesis testing procedures on the Examination for the Certificate of Proficiency in English (ECPE) data.
The ECPE data is collected by the English Language Institute of the University of Michigan, in which there are 2,922 test takers and 28 ECPE test items.
Three target attributes are considered, including morphosyntactic rules ($\alpha_1$), cohesive rules ($\alpha_2$) and lexical rules ($\alpha_3$).
The $Q$-matrix of the ECPE data is given in Table \ref{tab:Q-ECPE}.
Since the $Q$-matrix contains four identity submatrices, the testability conditions are satisfied.
A linear hierarchical structure $\mathcal{E}_0 = \{\alpha_3 \rightarrow \alpha_2 \rightarrow \alpha_1 \}$ is often considered in literature such as \cite{templin2014hierarchical}.

\begin{table}[h!]
  \centering
    \begin{tabular}{cccc}
    \toprule
    \multicolumn{1}{c}{\multirow{2}[4]{*}{Item}} & \multicolumn{3}{c}{Attributes} \\
\cmidrule{2-4}          & \multicolumn{1}{c}{Mor.rules ($\alpha_1$)} & \multicolumn{1}{c}{Coh.rules ($\alpha_2$)} & \multicolumn{1}{c}{Lex.rules ($\alpha_3$)} \\
    \midrule
    1     & 1     & 1     & 0 \\
    2     & 0     & 1     & 0 \\
    3     & 1     & 0     & 1 \\
    4     & 0     & 0     & 1 \\
    5     & 0     & 0     & 1 \\
    6     & 0     & 0     & 1 \\
    7     & 1     & 0     & 1 \\
    8     & 0     & 1     & 0 \\
    9     & 0     & 0     & 1 \\
    10    & 1     & 0     & 0 \\
    11    & 1     & 0     & 1 \\
    12    & 1     & 0     & 1 \\
    13    & 1     & 0     & 0 \\
    14    & 1     & 0     & 0 \\
    15    & 0     & 0     & 1 \\
    16    & 1     & 0     & 1 \\
    17    & 0     & 1     & 1 \\
    18    & 0     & 0     & 1 \\
    19    & 0     & 0     & 1 \\
    20    & 1     & 0     & 1 \\
    21    & 1     & 0     & 1 \\
    22    & 0     & 0     & 1 \\
    23    & 0     & 1     & 0 \\
    24    & 0     & 1     & 0 \\
    25    & 1     & 0     & 0 \\
    26    & 0     & 0     & 1 \\
    27    & 1     & 0     & 0 \\
    28    & 0     & 0     & 1 \\
    \bottomrule
    \end{tabular}%
   \caption{The $Q$-matrix for ECPE data. ``Mor." is short for ``morphosyntactic", ``Coh." is short for ``cohesive", and ``Lex." is short for ``lexical".}
  \label{tab:Q-ECPE}%
\end{table}%

Under the linear hierarchy $\mathcal{E}_0$, the   latent attribute profile set is $\mathcal{A}_0 = \{(0, 0, 0), (0, 0, 1), (0, 1, 1),$ $(1, 1, 1)\}$.
Under the null hypothesis, we fit a GDINA model with the profile set $\mathcal{A}_0$, and under the alternative, we fit a saturated GDINA model with all the possible attribute profiles.
We generated bootstrap samples  1000 times in parametric bootstrap and nonparametric bootstrap respectively.
The p-value obtained from parametric bootstrap is 0.041, while the p-value obtained from nonparametric bootstrap is 0.952.
Moreover, we also calculate the p-value corresponding to the na\"ive test using the conventional Chi-squared limiting distribution and get the p-value 0.02.
If we set the significance level to be 0.05, then by parametric bootstrap, we do reject the null hypothesis and conclude the linear hierarchy does not present in this data set; while if the significance level is set to be 0.01, we would not reject the null hypothesis and conclude there is such a linear attribute hierarchy.
This conclusion is consistent with that in \cite{templin2014hierarchical}.
For both significance levels, the nonparametric bootstrap does not reject the null hypothesis.

To conduct a more comprehensive study of the linear hierarchies among the three target attributes, we further test each linear hierarchy relationship separately and examine which one is the strongest. 
In particular, we consider the following various test settings:
\begin{itemize}
	\item $\text{H}_0: \mathcal{E}_0 = \{\alpha_3\rightarrow \alpha_2\}\ $ vs. $\ \text{H}_1$: no hierarchical structure $\mathcal{E}_1 = \emptyset$;
	\item $\text{H}_0: \mathcal{E}_0 = \{\alpha_2\rightarrow \alpha_1\}\ $ vs. $\ \text{H}_1$: no hierarchical structure $\mathcal{E}_1 = \emptyset$; 
	\item $\text{H}_0:\mathcal{E}_0 = \{\alpha_3\rightarrow \alpha_1\}\ $ vs. $\ \text{H}_1$: no hierarchical structure $\mathcal{E}_1 = \emptyset$;
	\item $\text{H}_0: \mathcal{E}_0 = \{\alpha_3\rightarrow \alpha_2 \rightarrow \alpha_1\}\ $ vs. $\  \text{H}_1: \ \mathcal{E}_1 =\{\alpha_3\rightarrow \alpha_2\}$;
	\item $\text{H}_0:\mathcal{E}_0 =  \{\alpha_3\rightarrow \alpha_2 \rightarrow \alpha_1\}\ $ vs. $\ \text{H}_1: \ \mathcal{E}_1 = \{\alpha_2\rightarrow \alpha_1\}$;
	\item $\text{H}_0: \mathcal{E}_0 = \{\alpha_3\rightarrow \alpha_2 \rightarrow \alpha_1\}\ $ vs. $\ \text{H}_1: \ \mathcal{E}_1 = \{\alpha_3\rightarrow \alpha_1\}$;
\end{itemize}
The resulting p-values for parametric bootstrap, nonparametric bootstrap and the na\"ive Chi-squared test under different settings are presented in Table \ref{tab:p-values}.
From the table, we can see that among all the settings, the p-values of nonparametric bootstrap are very large and therefore we do not reject the null hypotheses.
This is also consistent with our simulation study where we find nonparametric bootstrap is more conservative. 
The p-values of parametric bootstrap are much smaller than those of nonparametric bootstrap.
If we set the significance level to be 0.05, parametric bootstrap does not reject the null except for the settings ``$\{\alpha_3\rightarrow \alpha_2 \rightarrow \alpha_1\}$ vs.$\ \emptyset$" and ``$\{\alpha_3\rightarrow \alpha_2 \rightarrow \alpha_1\}$ vs. $\{\alpha_3\rightarrow \alpha_1\}$".
The p-values of the na\"ive Chi-squared test are of the similar scales of those in parametric bootstrap, and if the significance level is set to 0.05, the na\"ive Chi-squared test does not reject the null except for the settings ``$\{\alpha_3\rightarrow \alpha_2 \rightarrow \alpha_1\}$ vs. $\{\alpha_3\rightarrow \alpha_2\}$" and ``$\{\alpha_3\rightarrow \alpha_2 \rightarrow \alpha_1\}$ vs. $\emptyset$".
However, as we have shown in our simulation results, since the true limiting distribution of the LRT statistic is no longer the conventional Chi-squared distribution, the na\"ive test is not reliable.

\begin{table}[h!]
\centering
\begin{tabular}{cccc}
\toprule
Setting & Para-boot & Nonpara-boot & Chi-squared \\
\midrule
	$\{\alpha_3\rightarrow \alpha_2 \rightarrow \alpha_1\}$ vs. $\emptyset$  &     0.041                 &  0.952                       &          0.020         \\       
     $\{\alpha_3\rightarrow \alpha_2\}$ vs. $\emptyset$   &          0.052            &              0.511           &         0.072          \\
      $\{\alpha_2\rightarrow \alpha_1\}$ vs. $\emptyset$  &          0.057            &                0.722         &                   0.064 \\
      $\{\alpha_3\rightarrow \alpha_1\}$ vs. $\emptyset$  &          0.169            &              0.954           &                   0.066 \\ 
      $\{\alpha_3\rightarrow \alpha_2 \rightarrow \alpha_1\}$ vs. $\{\alpha_3\rightarrow \alpha_2\}$  &   0.098                   &             0.962            &          0.047         \\   
      $\{\alpha_3\rightarrow \alpha_2 \rightarrow \alpha_1\}$ vs. $\{\alpha_2\rightarrow \alpha_1\}$  &  0.073                    &        0.906                 &       0.052            \\     
      $\{\alpha_3\rightarrow \alpha_2 \rightarrow \alpha_1\}$ vs. $\{\alpha_3\rightarrow \alpha_1\}$  &   0.026                   &             0.625            &       0.051            \\       
      
\bottomrule        
\end{tabular}
\caption{p-values for different testing settings.}
\label{tab:p-values}
\end{table}

For testing a single linear hierarchy relationship versus none hierarchical structure (i.e., the second to the fourth tests in Table \ref{tab:p-values}), since the alternative hypothesis is the same while the null hypothesis varies, a larger p-value suggests a stronger pre-requisite relationship.
Therefore based on the results in Table \ref{tab:p-values} that  the p-value for ``$\{\alpha_3 \rightarrow \alpha_1\}$ vs. $\emptyset$" is the largest among these three tests, we can see that the prerequisite relationship between the third attribute and the first attribute is the strongest.
Similarly, for testing the whole linear hierarchical $\mathcal{E}_0$ versus a single linear hierarchy (i.e., the last three tests in Table \ref{tab:p-values}), since now the null hypothesis is the same and the alternative hypothesis varies, a smaller p-value indicates a stronger pre-requisite relationship.
In this case, the prerequisite relationship between the third attribute and the first attribute seems still the strongest.

\section{Discussions}
\label{sec-discuss}

In this paper, we consider the hypothesis testing problem for latent hierarchical structures in CDMs. 
We first discuss  the testability issues and present sufficient conditions for testability.
Under the testability conditions, we study asymptotic properties of the likelihood ratio test and show the practical difficulties of directly using the limiting distribution of the LRT statistic to test latent hierarchies.
We then compare two resampling-based testing procedures including parametric bootstrap and nonparametric bootstrap through comprehensive simulations under different settings and recommend using parametric bootstrap for testing latent hierarchical structures.

In this paper, we mainly focus on the hypothesis testing where the hierarchical structure is fully specified and all the latent attribute profiles that respect the hierarchy exist in the population. 
In many applications, the number of  latent attributes $K$  could be large, leading to a high-dimensional space for all the possible configurations of the attributes,   where the number of potential attribute profiles can be even larger than the sample size. For scientific interpretability and practical use, it is often assumed that not all the possible attribute profiles exist in the population. In such cases, to test hierarchical structures, we may  perform the  selection of significant latent attribute profiles first and then conduct testing procedures.

This paper proposes to use parametric bootstrap, which is a resampling-based procedure and can be computationally expensive, especially for large-scale data sets.
Therefore it would be useful to develop more efficient testing procedures.
Moreover, further theoretical results are needed to better characterize the asymptotic distribution of the likelihood ratio test with the presence of latent variables and complex constraint structures as in CDMs.

\paragraph{Supplementary Material}
More comprehensive simulation results are presented in the supplementary material.
Specifically, bootstrap results for DINA and GDINA models under both null hypothesis and alternative hypothesis with different sample sizes and noise levels are plotted there.
We also include the codes for simulations and real data analysis.

\paragraph{Acknowledgement}
This research is partially supported by National Science Foundation CAREER SES-1846747 and Institute of
Education Sciences R305D200015. The authors thank Dr. Yuqi Gu for helpful discussions.

\small
\bibliographystyle{apalike}
\bibliography{bibref.bib}

\newpage
\appendix

\section*{Supplementary Material: Additional Simulation Results}

\begin{figure}[H]
    \centering
    \subfigure{
    \includegraphics[width=7in]{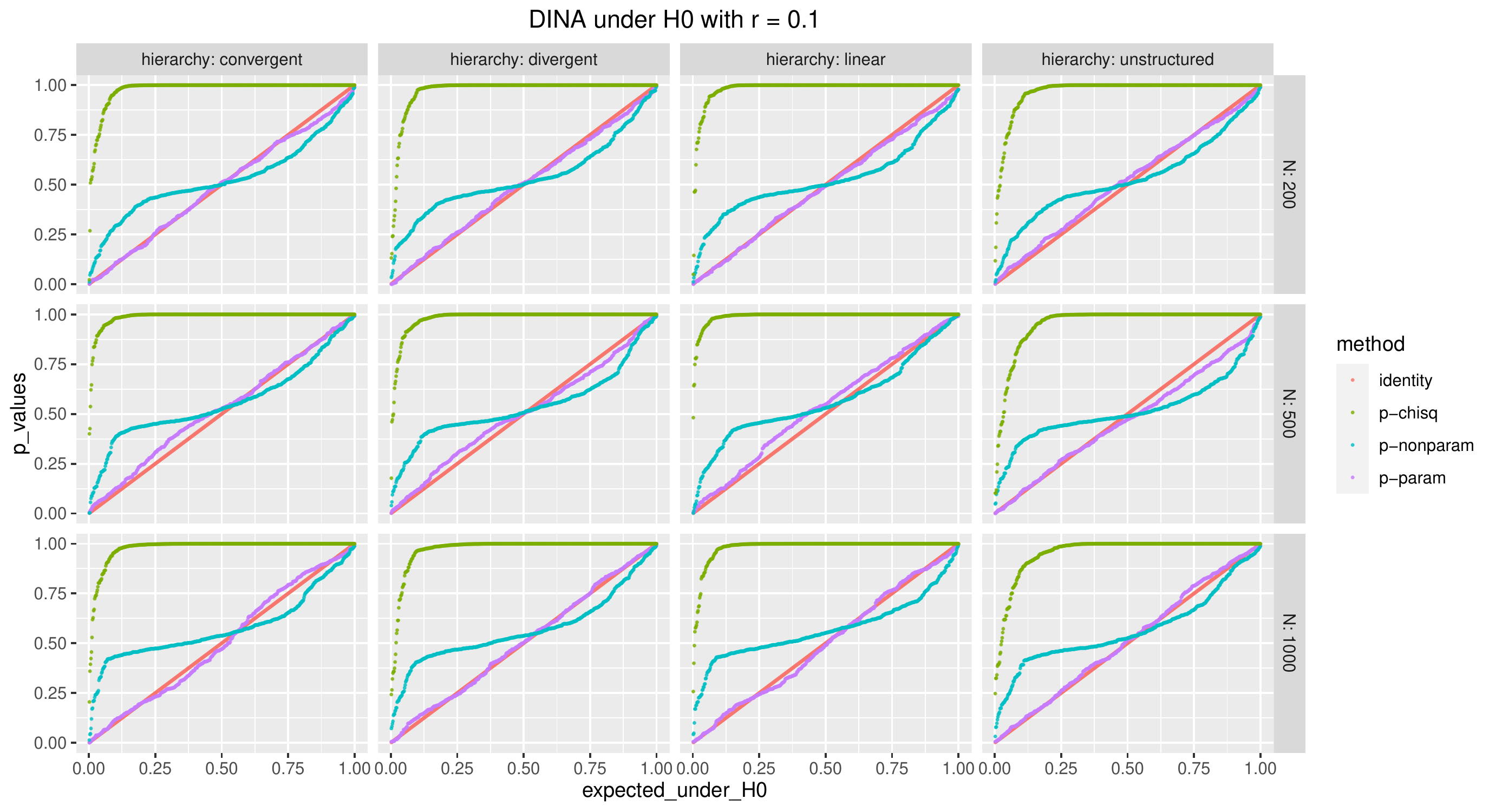}
    }
    \subfigure{
    \includegraphics[width=7in]{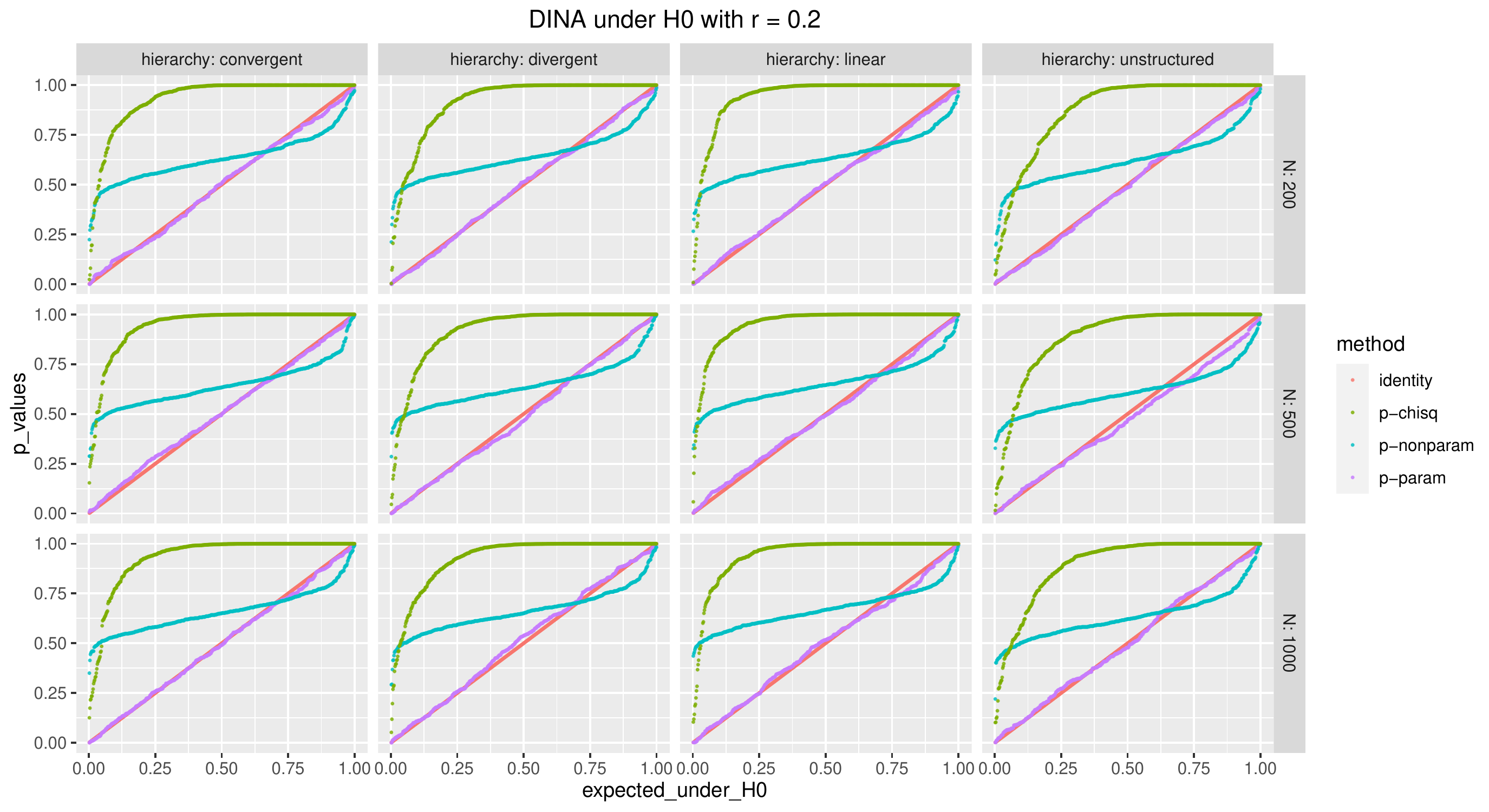}
    }

\caption{Bootstrap results for DINA under null hypothesis}
    \label{fig:dina-qq-full}
\end{figure}{}

\begin{figure}[H]
    \centering
    \subfigure{
    \includegraphics[width=7in]{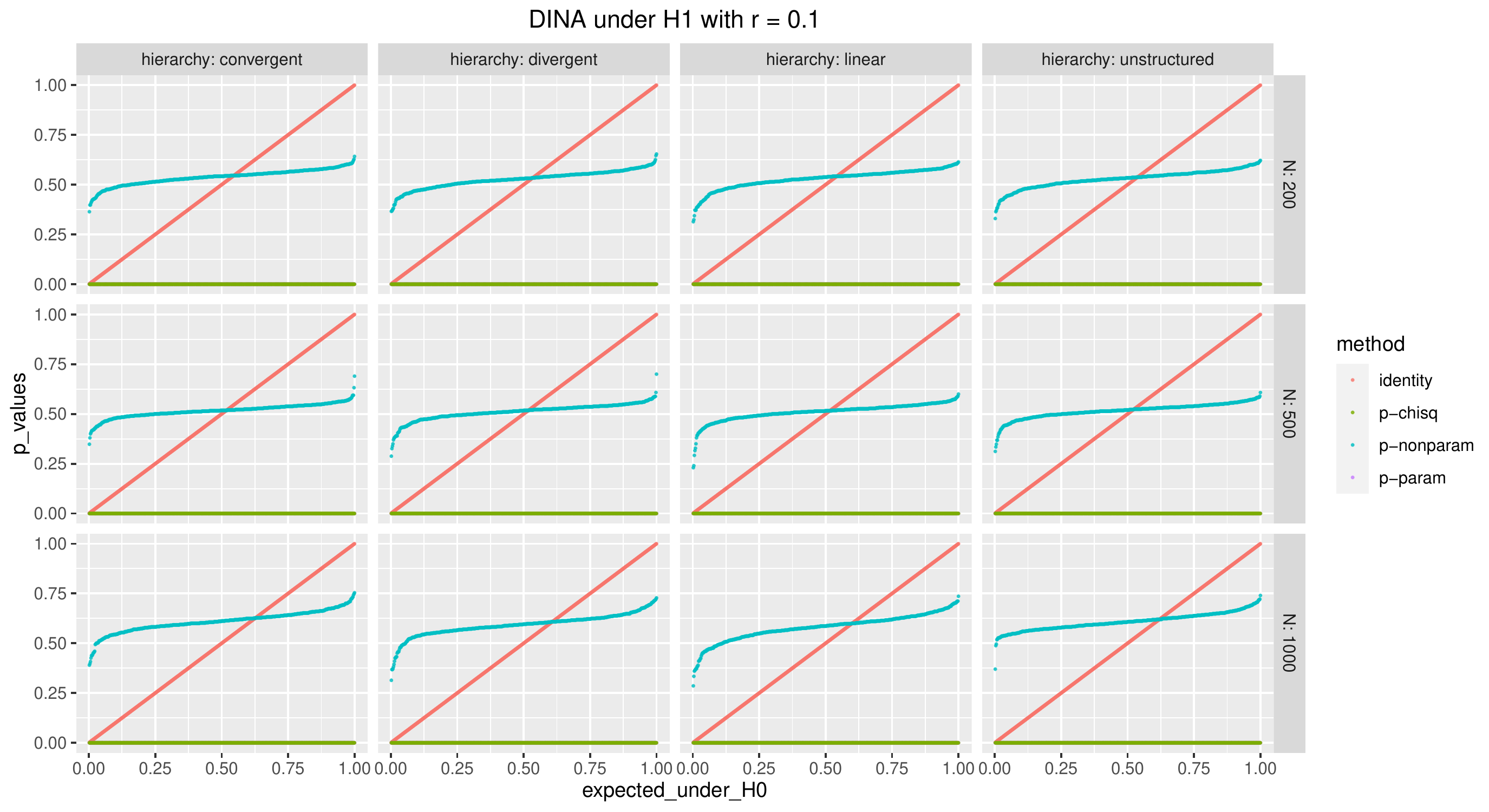}
    }
    \subfigure{
    \includegraphics[width=7in]{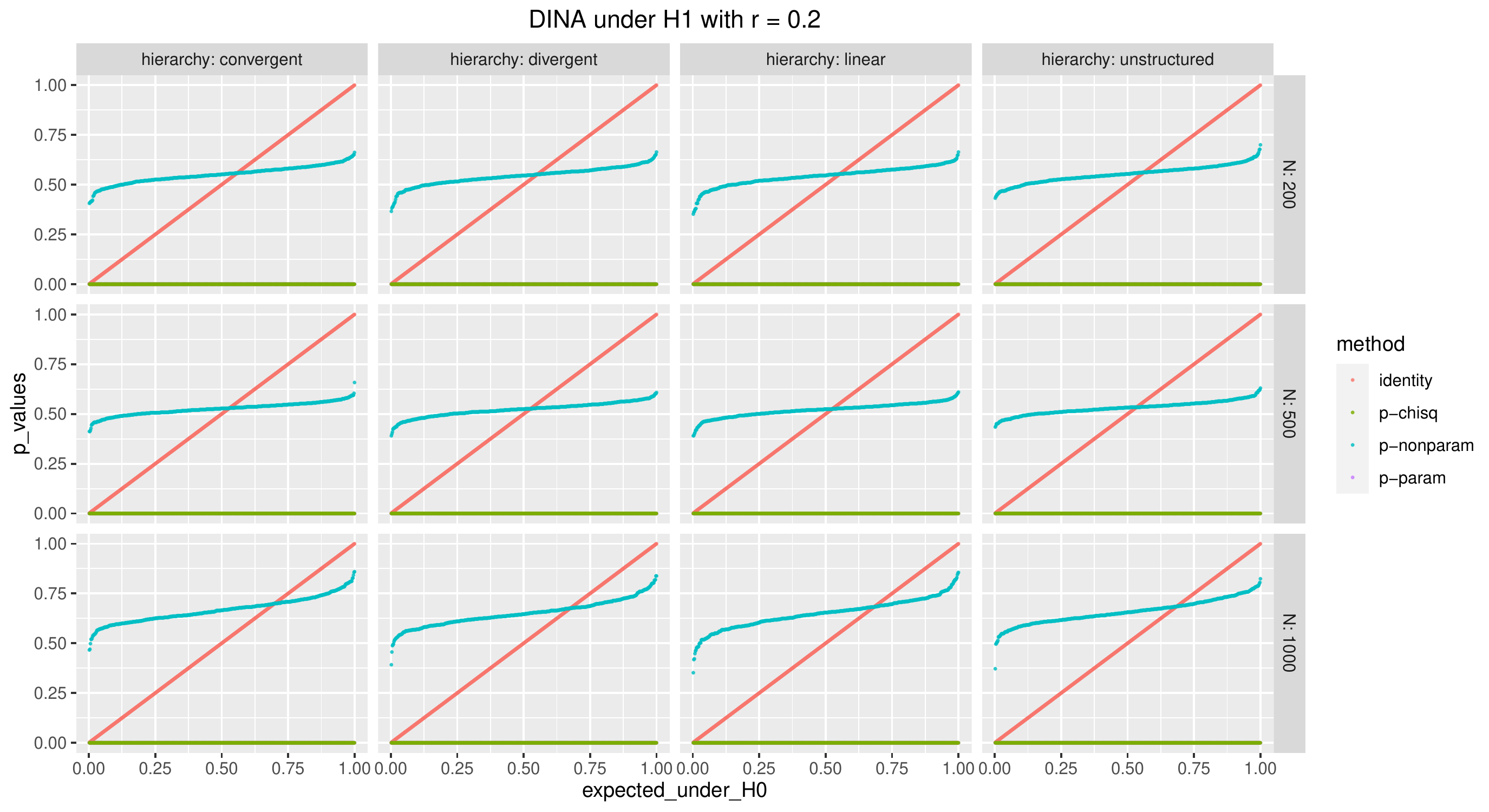}
    }
	\caption{Bootstrap results for DINA under alternative hypothesis}
    \label{fig:dina-qq-power-full}
\end{figure}{}

\begin{figure}[H]
    \centering
    \subfigure{
    \includegraphics[width=7in]{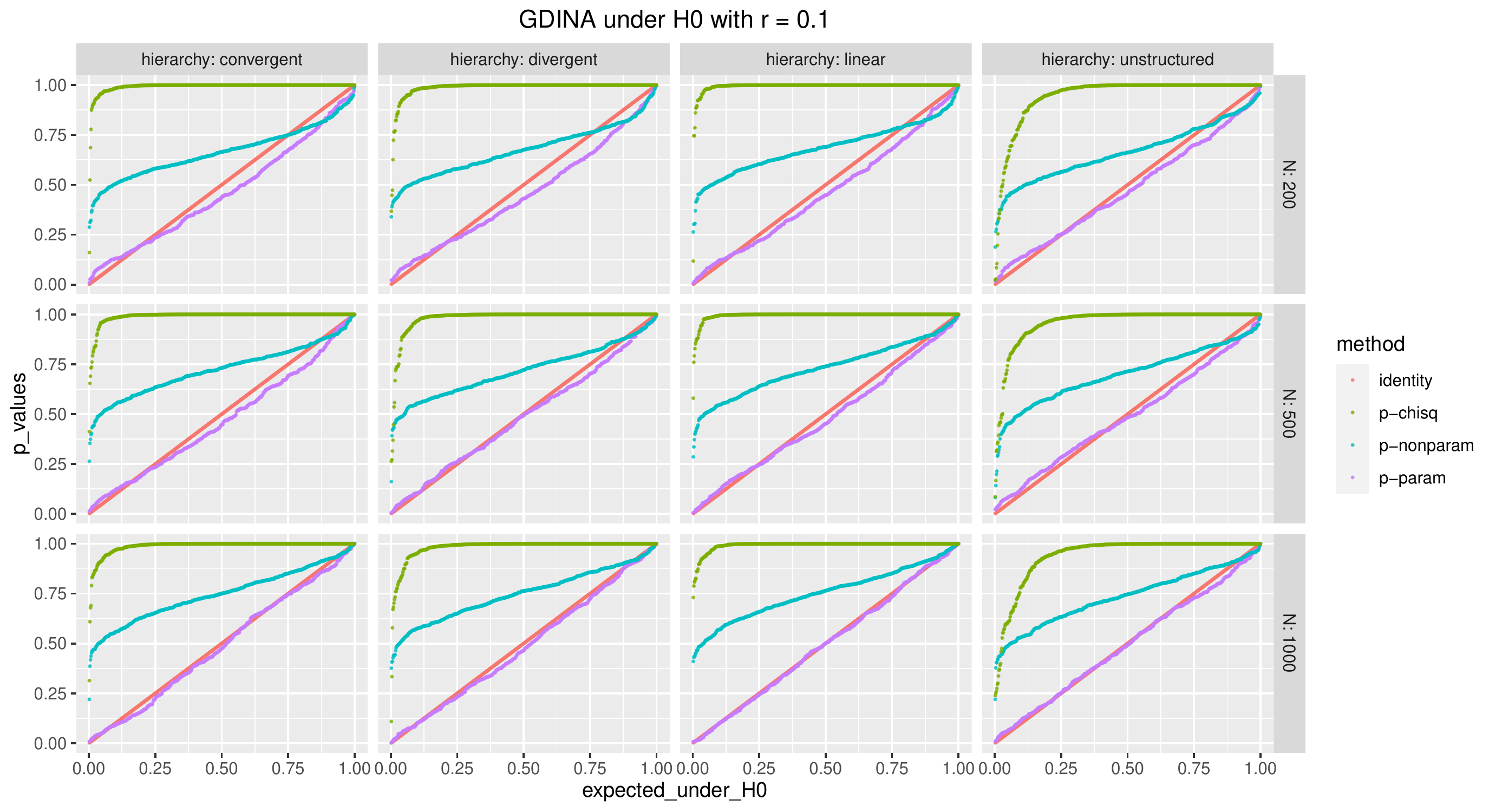}
    }
    \subfigure{
    \includegraphics[width=7in]{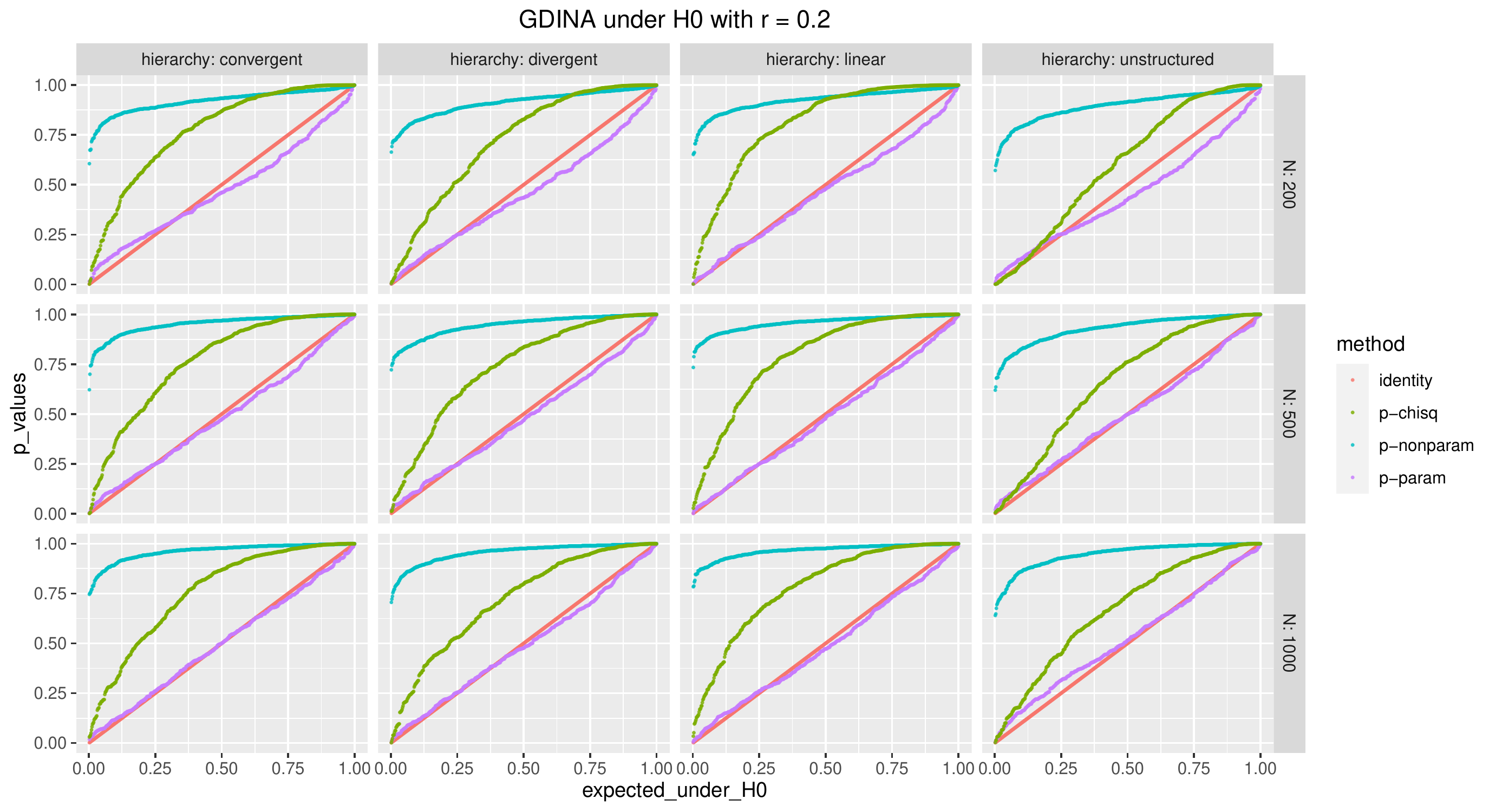}
    }
	\caption{Bootstrap results for GDINA under null hypothesis}
    \label{fig:gdina-qq-full}
\end{figure}{}

\begin{figure}[H]
    \centering
    \subfigure{
    \includegraphics[width=7in]{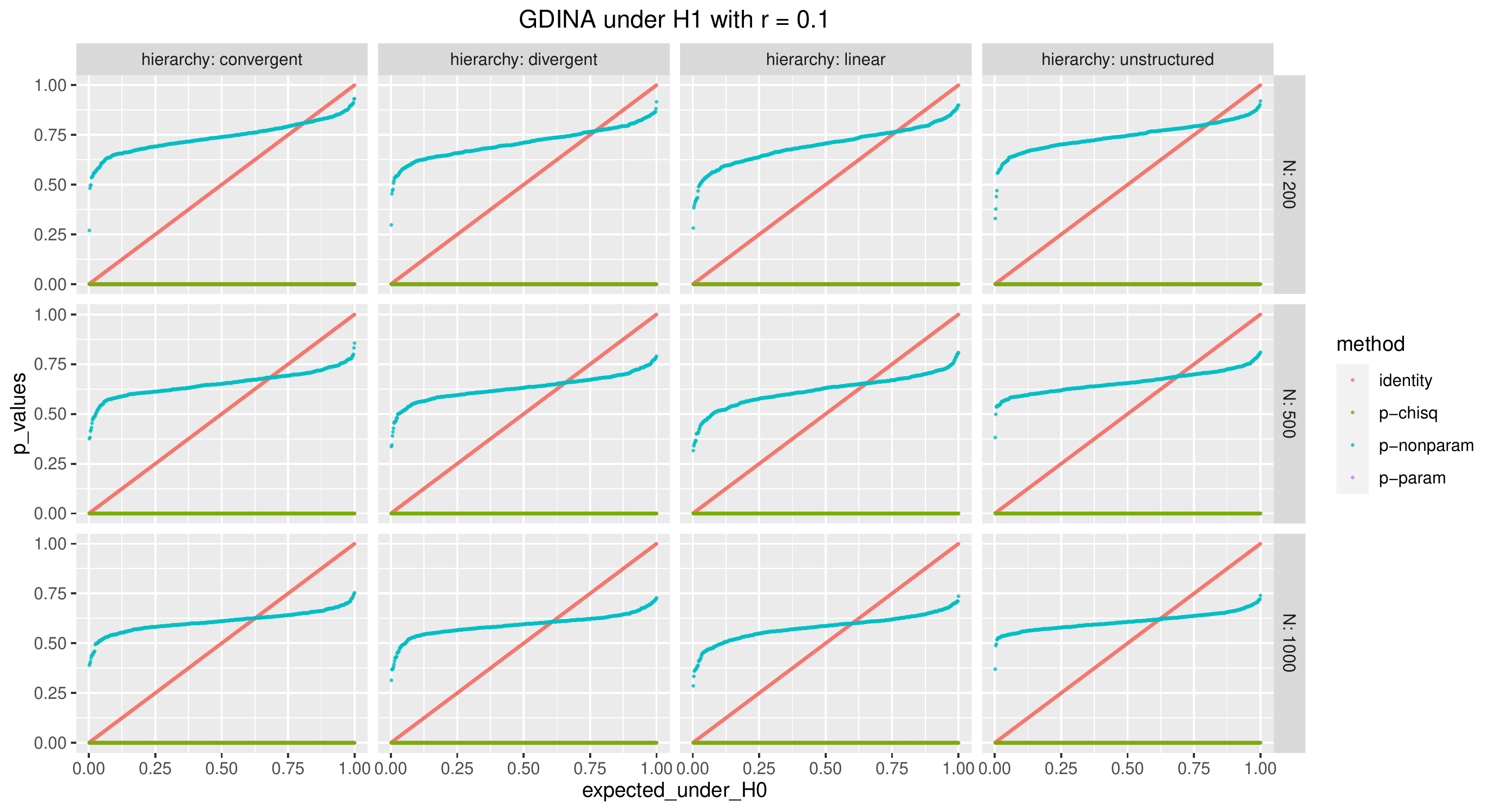}
    }
    \subfigure{
    \includegraphics[width=7in]{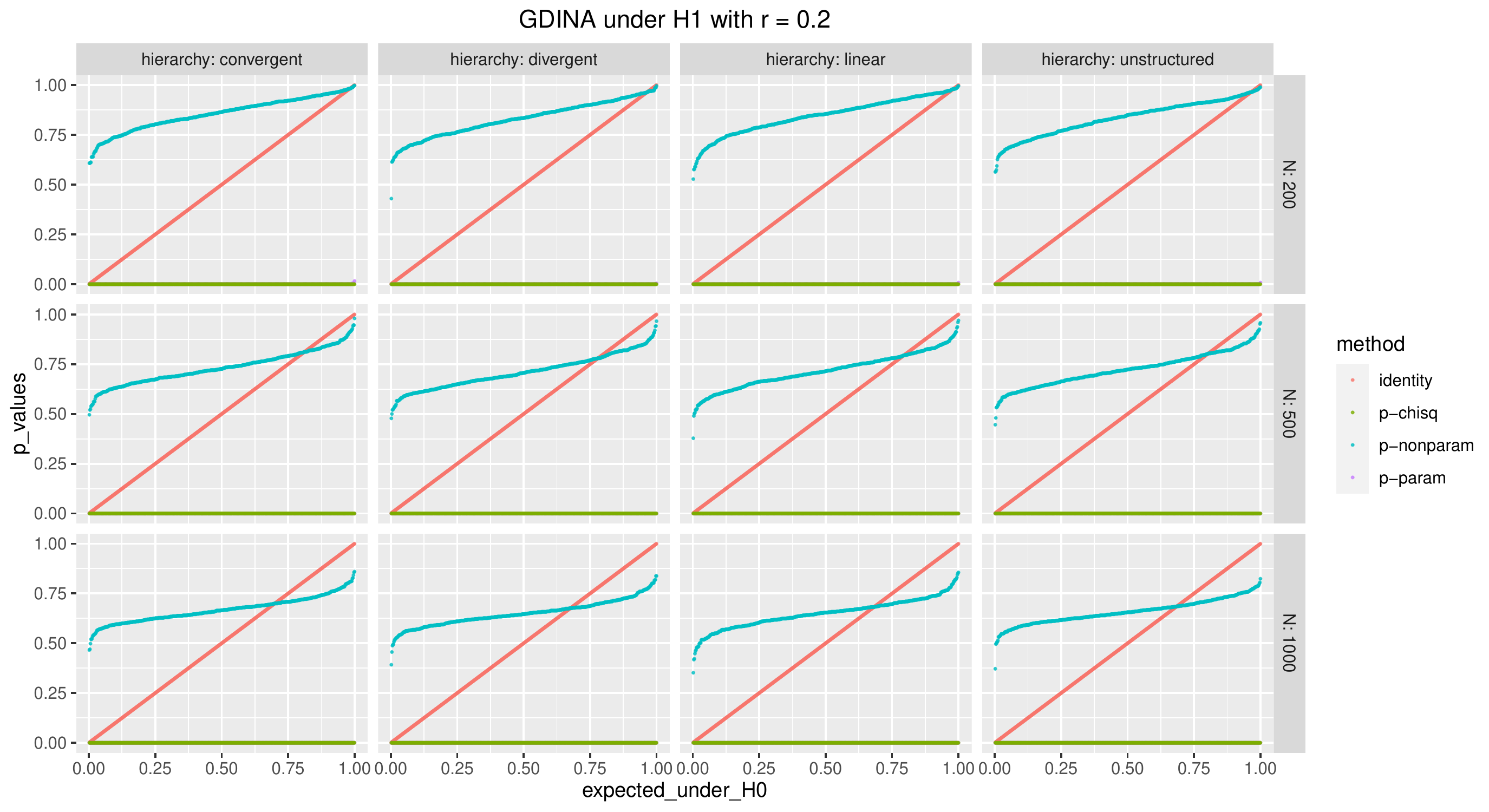}
    }
	\caption{Bootstrap results for GDINA under alternative hypothesis}
    \label{fig:gdina-qq-power-full}
\end{figure}{}

\end{document}